# Probing Molecular Ordering in the Nematic Phases of *para-Linked* Bimesogen Dimers through NMR Studies of Flexible Prochiral Solutes

Leah M. Heist[1], Edward T. Samulski[1], Chris Welch[2], Ziauddin Ahmed[2], Georg H. Mehl[2], Alexandros G. Vanakaras[3] and Demetri J. Photinos[3]

[1] *Department of Chemistry, University of North Carolina, USA*
[2] *Department of Chemistry, University of Hull, Hull, UK*
[3] *Department of Materials Science, University of Patras, Patras, 26504, Greece*

**Abstract.**

The quadrupolar splittings of perdeuteriated n-decane dissolved in nematic phases formed by mesogenic dimers of the CBnCB series, for n=7,9,10,11 are measured throughout the entire temperature range of these phases. The results of the measurements, are reported together with related measurements using the common nematic phase of 5CB as a solvent for n-decane. The data obtained from the $^{13}$C spectra of the cyanobiphenyl mesogenic units of the monomeric and dimeric solvent molecules yield the order parameter of those units. The information obtained from this set of experiments is used to elucidate the structure of the low temperature ($N_X$) and the high temperature (N) nematic phases of CBnCB dimers with n=7,9,11. The polar twisted nematic ($N_{PT}$) model is found to provide a consistent description not only of these measurements, but also of NMR measurements previously reported in the literature for these phases. These findings suggest that the high temperature nematic (N) is not a common, locally uniaxial and apolar nematic, but rather a nematic phase consisting of $N_{PT}$ clusters. The twist-bend ($N_{TB}$) model, often identified with the $N_X$ phase, is shown to be inadequate to account even qualitatively for crucial features of the experimental findings.

## 1. Introduction.

The definitive characterization[1–4] of a second, lower temperature, nematic in the phase sequence of the symmetric, odd-spacer length, cyano-biphenyl dimers CBnCB, terminated a two decade uncertainty as to the nature of this phase. At the same time, this dilemma opened up what is presently the most vigorously researched topic in liquid crystal science: understanding the structure[5–17] and potential applications [18–22] of the second, lower temperature nematic phase, hereafter referred to as $N_X$. The phase was termed nematic because no positional ordering of the molecules could be detected experimentally [1,2]. However, the $N_X$ phase in the odd-spacer length CBnCB dimers shows a distinct difference from its higher temperature nematic phase, a macroscopically uniaxial and apolar phase hereafter denoted by N for brevity. The $N_X$ phase exhibits a ~10 nm spatial modulation of the orientational ordering that persists over domain sizes exceeding the μm range. The structure of the $N_X$ phase remains debated because of an identification [23] with the twist-bend nematic phase ($N_{TB}$). Subsequent accumulation of experimental observations revealing inconsistencies of the $N_X$ properties with those implied by the $N_{TB}$, such as the length scale of the spatial modulation and the symmetry of the local molecular ordering, led to the formulation [24,25] of the polar twisted nematic ($N_{PT}$) model. The latter has been successful in accounting for the main structural feature of the Nx phase, namely short-pitch modulation of the orientational ordering in the presence of complete positional disorder together with fundamental features of the molecular interactions as revealed by NMR spectroscopy. [25,26]

A distinguishing NMR signature of the Nx phase, which has been used to clearly identify the N-$N_X$ phase transition is the doubling of spectral lines associated with pairs of prochiral sites (enantiotopic discrimination). [2,4,7,8] Such pairs show coincident residual dipolar and quadrupolar NMR spectral lines in the N phase. The appearance of enantiotopic discrimination indicates the loss of equivalence regarding the orientational ordering of the two sites that form the prochiral pair. [27–30] The ordering of small rigid solutes in both nematic phases exhibited by dimers of the CBnCB series has been studied experimentally [2,31] by NMR and the results were analysed consistently within the $N_{PT}$ model.[26] It was found that three distinct mechanisms could in principle underlie the appearance of enantiotopic discrimination in the $N_X$ phase. The primary mechanism, in the case of small solute molecules, is based on the polar orientational ordering of the solvent phase. This mechanism was shown to be incompatible with the $N_{TB}$ model. A second mechanism, applicable to larger solute molecules in addition to the first mechanism, results from the modulation of the orientational ordering, whereby different parts of a given solute molecule sample different ordering environments within the short-pitch (~10 nm) solvent phase. A third mechanism reflects directly the chiral asymmetry imposed on the dimer solvent molecules as a result of the modulation of the ordering: the dimers of the CBnCB series are intrinsically achiral and flexible whilst the modulation of the orientational ordering produces thermodynamically equivalent, short pitch, twisted domains of opposite handedness. Within each such domain, the conformational statistics of the dimer molecules are chirally biased, in the sense that a probability imbalance between enantiomer conformations is introduced, and thus the statistical achirality of these molecules is lost. In the $N_X$ phase, this third mechanism is expected to have marginal contributions due to the estimated smallness of the induced chiral imbalance on the (intrinsically achiral) solvent molecules. [24–26]

The $N_{PT}$ model originated from a molecular theory [24] of the positionally disordered fluid phases that can be formed by idealized mesogenic dimers. Three such phases were obtained: an isotropic fluid at high temperatures, a common, uniaxial apolar nematic (Nu) at intermediate temperatures, below which a polar and twisted ($N_{PT}$) phase appears. Briefly, the $N_{PT}$ phase has no axis of full rotational symmetry but only a polar director $\hat{m}$, which is a local $C_2$ axis undergoing tight twisting at constant pitch *p*, on the order of the dimer molecular length, while remaining perpendicular to a fixed direction, the helix axis $\hat{n}_h$. It is noted that the concept allows for a pitch length change with temperature. Regarding molecular structures, the basic requirement for the formation of the $N_{PT}$ phase is molecular polarity. Such polarity typically originates from the molecular shape ("steric dipole"), most commonly encountered in V-shaped molecules, rigid or flexible. In the latter case, the statistically achiral molecules can acquire a net statistical twisting, induced by their tightly twisted environment within a domain of given handedness. Left- and right-handed twisting of $\hat{m}$ are thermodynamically equiprobable in an unbiased sample. Accordingly, such a sample is overall apolar and achiral and also macroscopically uniaxial. The second rank ordering tensors of any molecular segment have $\hat{m}$ as a common principal axis while the inversion $\hat{m} \Rightarrow -\hat{m}$ is not a local symmetry operation.

The purpose of this work is to probe the molecular structure in the two nematic phases of odd-spacer CBnCB dimers by means of NMR experiments on elongated flexible solute molecules. The elongated and flexible solutes may span a finite portion of the tight pitch in the $N_{PT}$ and are thus expected to provide information on the second mechanism described above, interactions emerging from the spatial modulation of the ordering in the $N_X$ phase. The flexibility of the solute molecules can also provide information on the extent of induced chirality imbalance in the twisted domains. Fully deuteriated $n$-alkanes were chosen as advantageous solutes for this study due to their simple structure, the ease of selecting the appropriate molecular elongation, and their very successful use and modelling as solutes in common nematic solvents. [32–34] The experimental results on $n$-decane are presented and analysed in the context of the $N_{PT}$ model. The analysis is extended to existing NMR measurements of CBnCB dimers in their nematic phases.[2,35] A coherent description of such experimental results is obtained within the $N_{PT}$ picture of dimer organization in this tight-pitch $N_X$ phase.

Elucidating the molecular organization in the high temperature N phase formed by dimers which exhibit an $N_X$ phase is crucial to understanding the nature of the N-$N_X$ transition and eventually the $N_X$ phase itself. To that end, the same NMR techniques and analysis of the spectra used for the $N_X$ phase are also applied to the N phase of dimers, where no enantiotopic discrimination is observed. Two possible types of phase structure can be envisaged for this N phase:

1. a locally uniaxial structure, described by a nematic director **n** that constitutes an axis of full rotational symmetry, as in "conventional" nematics which show no enantiotopic discrimination. This is the picture routinely adopted for the high temperature N phase, albeit without any definitive evidence. The N-$N_X$ phase transition in this case entails a change of local symmetry.
2. a phase consisting of clusters having a polar twisted array of dimers—identical to the structure in the domains of the low temperature, $N_X$, phase (i.e. locally polar, with a tightly twisting polar director), except that the spatial extent of these clusters in the N phase is much smaller than the domains in the $N_X$ phase. In short, the N phase in this picture consists of small $N_X$ clusters of opposite handedness. The rapid diffusion of the solute molecules among clusters of opposite handedness would wipe out any enantiotopic discrimination. And, the N-$N_X$ transition in this case signals the coalescing of the clusters to form macroscopic domains of the same symmetry.

A previous analysis, based on the $N_{PT}$ model, concluded that the NMR spectra obtained from small rigid solutes [26] would be compatible with either of the above two alternatives for the high temperature N phase. However, as shown in the present study, the detailed analysis of NMR spectra obtained from alkane solutes does differentiate between the two alternatives and, in fact, such differentiation favours the second alternative.

The remainder of the paper is organized as follows. The next section contains the description of the NMR experiments and the presentation of the spectroscopic results. Section 3 relates the experimentally measured spectra to the orientational order in the solvent phase, in terms of the solvent-solute intermolecular interactions imbedded in the potential of mean torque (PMT). In section 4, the calculated spectra according to the $N_{PT}$ model are compared to the experimental NMR measurements described in section 2 as well as to measurements that have appeared in the literature on CBnCB dipolar couplings, both in the $N_X$ and its high temperature counterpart N. The results are discussed and the conclusions are drawn in section 5. Chemical structures and phase transition temperatures of relevant members of the CBnCB series are given in the appendix. The full synthetic details have been reported elsewhere. [36] Experimental details and a collection of NMR data are included in a Supplementary Information File.

## 2. $^2$H and $^{13}$C NMR spectra in the nematic phases of CBnCB dimers.

The order parameters of the cyano group, i.e., the orientation of the –CN bond (para axis) of cyano-biphenyl terminated dimers with alkyl spacers with 7, 9, 10 and 11 methylene carbons, CB7CB, CB9CB, CB10CB and CB11CB respectively, and the ordering of the solute "probe" decane-$d_{22}$ have been determined using carbon and deuterium NMR, respectively. The observations are plotted versus reduced temperature; the CBnCB dimer transitions are shown in Appendix 1.

***Deuterium NMR:*** The 2 wt % decane-$d_{22}$ in CBnCB and 5CB samples were prepared in 5mm O.D. NMR tubes and degassed using three freeze-pump-thaw cycles. Spectra were taken with a Bruker DMX360 FT NMR spectrometer with a modified probe to access the transition temperatures. $^2$H NMR spectra were recorded at 55.3 MHz (Bruker DMX360 FT NMR spectrometer). The sample was mixed in isotropic phase, introduced into the spectrometer, and after equilibration to the desired temperature, 5000 scans were averaged with a spectral width of 100 kHz. A quad echo pulse sequence was used with a 90° pulse width of 2 μs (5 dB), 100 ms recycle time, and 18 and 10 μs between the first and second 90° pulse respectively.

Figure 1 summarizes the $^2$H NMR spectra of the decane-$d_{22}$ solute probe in CBnCB dimer solvents (n = 7, 9, 10, 11) and the nematic 5CB solvent versus respective reduced temperatures. For simplicity, only the quadrupole splitting, $\delta\nu$, of the methylene group adjacent to the methyl—the deuterons on the penultimate "C2" carbons of decane—are plotted. (The experimental spectra and tabulated splittings values are presented in the Supplementary Information.) As the temperature decreases into the respective Nx phases, there is an obvious doubling of the C2 quadrupolar splittings for CB9CB and CB7CB; the splitting is not readily apparent in CB11CB (it is too small). The doubling of the C2 splitting is absent in the nematic phase of the even dimer CB10CB and the monomeric solvent 5CB.

The magnitude of decane probe ordering in CB10CB is significantly higher than that in the odd dimers and 5CB solvents suggesting a qualitatively different type of organization and/or interactions in the N phase for the even dimer. For the odd dimers the ordering of the mesogenic cores—the order parameter of the *para*-axes of the cyanobiphenyl units—is explicitly addressed by measuring the chemical shift anisotropy of $^{13}$C NMR resonance of the –CN groups (see SI). The order parameters $S_{CN}$ initially increase steadily in the nematic phase and then begin to decrease as the Nx phase is approached for CB9CB, CB7CB; the temperature dependence of the monomer solvent 5CB is typical (See Fig. 2).

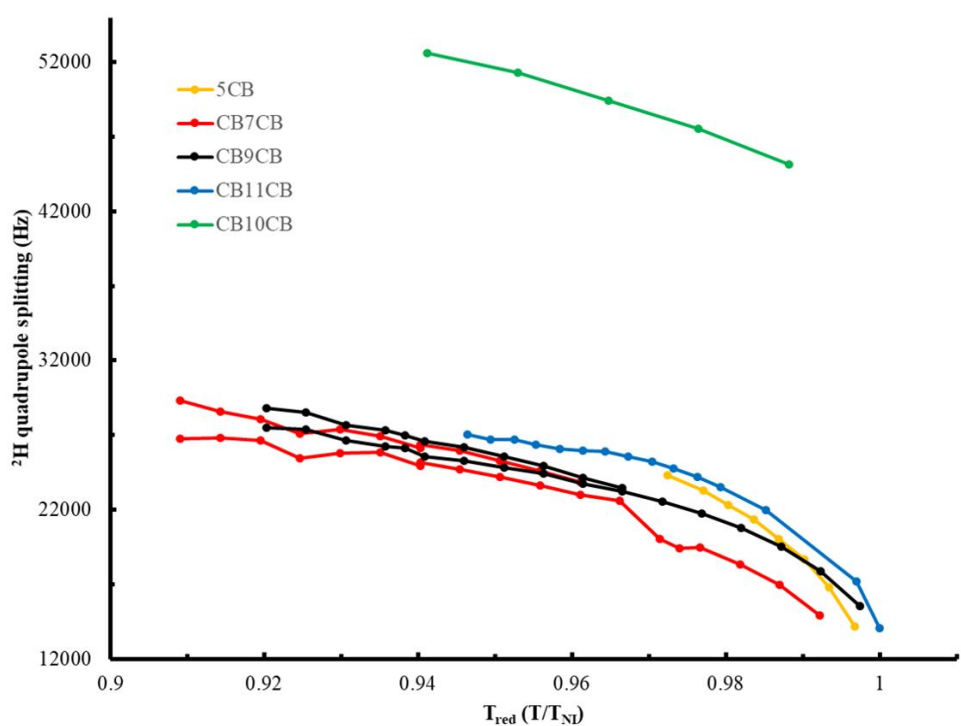

**Figure 1.** The reduced temperature dependence of the quadrupole splittings, $\delta\nu$, of the penultimate carbon methylene of decane-d$_{22}$ probe solute dissolved in CBnCB dimers, with n = 7, 9, 10, 11, and 5CB monomer solvents.

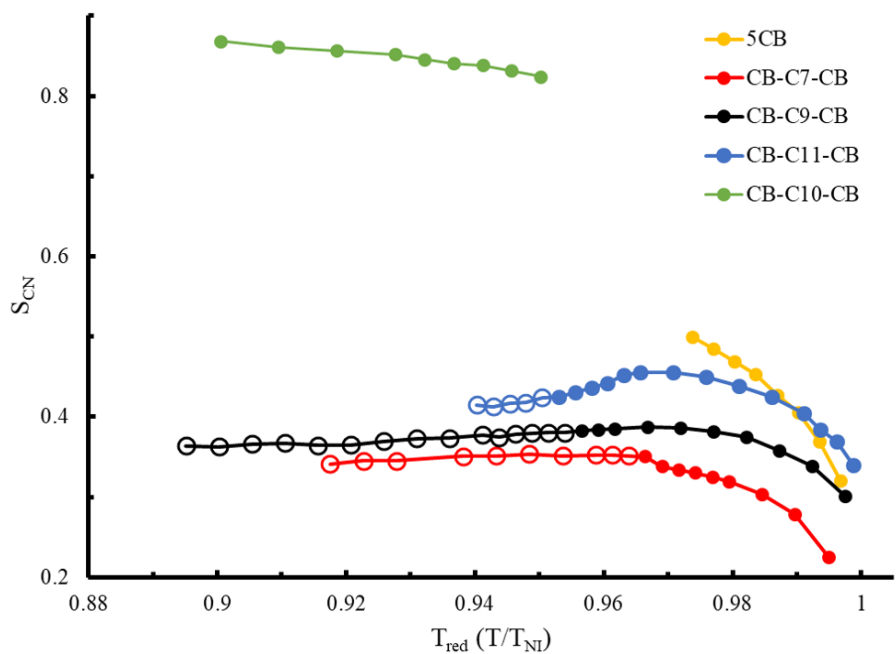

**Figure 2.** The (reduced) temperature dependence of the order parameter of CN unit, $S_{CN}$ is shown for the dimers CB11B, CB9CB, CB7CB, and the monomer 5CB. For the odd-spacer dimers the open circles indicate the Nx phase as determined by DSC experiments.

Note that the CN order paramater shows a clear maximum at a reduced temperature which falls well within the N phase of CB11CB, beyond which the order parameter decreases with decreasing temperature both in the remainder of the N phase range and in the $N_X$ phase. An anlogous trend, albeit weaker, is observed in the temperature dependence of the order paramater of CB9CB. Similar nonmonotonic temperature dependences have been reported [37,38,14,39] for other dimer compounds exhibiting the $N_X$ phase and suggest that the high temperature N phase of all these compounds is not a common (locally uniaxial, apolar, achiral) phase. We shall return to this point in more detail in sections 4 and 5.

As indicated in figure 2, the degree of ordering and its temperature dependence differs for the four LC compounds shown. These differences are reflected in the quadrupolar spectra obtained from decane dissolved the respective compounds. To detect a possible solvent-dependence in the relative ordering of the decane segments we have scaled the methylene splittings with respect to the methyl splittings. This eliminates the differences in the extent of overall ordering in the different solvents and allows a comparison of the relative ordering experienced by the different methylene segments of the decane solute and for different solvents. The scaled splittings are plotted in figure 3, as functions of the respective reduced temperatures, for the measured spectra of decane dissolved in the nematic phases of the CBnCB dimers with n=7,9,10 and also of 5CB. It is apparent from this figure that the reduced plots for the odd-n dimers and for 5CB are very close to having a common temperature dependence. This would suggest a common ordering mechanism of the decane chain solute in the different solvents. As the common element in these solvents is the mesogenic unit, a possible interpretation would be that (i) the decane solute is primarily ordered by interactions with the mesogenic units and (ii) the pendant chain in 5CB and the spacer chain of the odd-dimers either behave similarly, regarding their effects on the ordering of the decane solute, or these effects are marginal. On the other hand, the deviations of the scaled splittings obtained in the even dimer CB10CB solvent, mainly reflecting the elevated ordering of the terminal parts of decane (methyl group and adjacent methylene) relative to the central parts (methylenes 3,4,5), indicate possible differences in the ordering mechanism in the CB10CB solvent compared to the odd-spacer dimers. As will be discussed below, this could be related to the different organizational structure within the nematic phase formed by CB10CB and <u>both</u> nematic phases (the low temperature Nx as well as the high temperature N) formed by CB7CB and CB9CB.

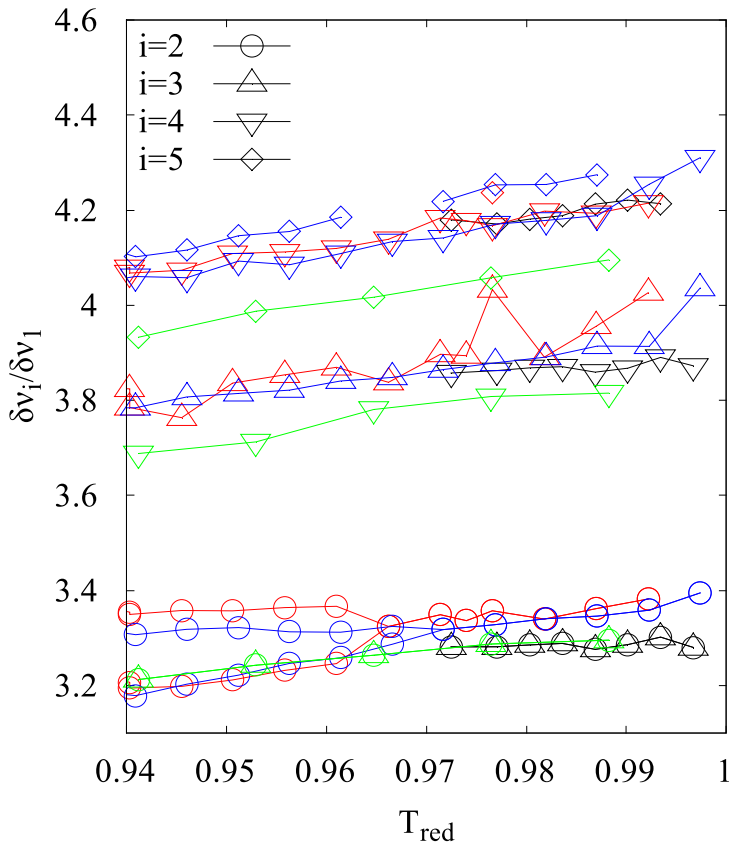

**Figure 3.** Plots of scaled quadrupolar methylene splittings of decane, given by $\delta v_i / \delta v_1$, with $i = 2, 3, 4, 5$ according to the site numbering where $i = 1$ corresponds to the methyl group, $i = 2$ is the penultimate carbon, etc. (See Fig. 4). The splittings are measured for deuterated decane dissolved in the nematic solvents 5CB (black), CB7CB (red), CB9CB (blue) and CB10CB (green). $T_{red} = T/T_{N-I}$ denotes the reduced temperature in each solvent.

## 3. Relation of NMR Spectra to Molecular Ordering for Flexible Solutes in the Polar-Twisted Phase.

The NMR spectra analysed in this work involve deuterium residual quadrupolar couplings, $^{13}$C chemical shift anisotropy and residual dipolar couplings. The spectroscopic observable [26,30] obtained from a deuterated site *s* belonging to a solute molecule is the residual quadrupolar splitting $\delta v_s$. In the case, typical of quadrupolar spectra obtained in nematic media, where the deuterated site undergoes rapid reorientational motion on the time scale of the NMR measurement, the quadrupolar splitting is given by

$$\delta v_s = \frac{3}{2} q_s \left\langle P_2(\hat{H} \cdot \hat{e}_s) \right\rangle \quad , \tag{1}$$

where $P_2$ denotes the second Legendre polynomial, $\hat{e}_s$ is the direction of the principal axis of the electric field gradient tensor at the position of the quadrupole moment of the deuterium nucleus, $\hat{H}$ is the direction of the spectrometer magnetic field, and $q_s$ is the quadrupolar coupling constant. For deuterons bonded to aliphatic carbons $q_{s=CD} \simeq 167\,\text{kHz}$. The angular brackets indicate thermal averaging over the orientations of the deuterated site. When the site belongs to a flexible molecule, such orientational averaging is the combined outcome of the reorientations of the entire molecule and its conformational changes that modify the orientation of the deuterated site relative to the direction $\hat{H}$ of the field. In a spatially modulated phase, as is the case of the N$_{PT}$, where the polar axis twists along (remaining normal to) the helical axis (hereafter identified with the Z macroscopic direction), the averaging is also influenced by the changes of the position of the deuterated site during the NMR time-scale. Thus, for a flexible solute molecule that (i) undergoes rapid reorientations, (ii) visits rapidly all the accessible conformations and (iii) diffuses rapidly, traveling several pitch lengths (*h*) within a domain of specific handedness in the N$_{PT}$ phase, the averaging indicated by the angular brackets in eq(1) for any molecular quantity $Q$ is expressed as

$$\left\langle Q \right\rangle \equiv \frac{1}{\zeta h} \int_0^h dZ \sum_n e^{-E_n/kT} \int d\omega\, Q_n(\omega, Z) e^{-V_n(\omega, Z)} \quad , \tag{2}$$

with $\zeta \equiv \sum_n e^{-E_n/kT} \int d\omega\, e^{-V_n(\omega, Z)}$.

Here $Q_n(\omega, Z)$ is the value of the quantity being averaged when the molecule is in its *n-th* conformation, has orientation $\omega$ and position $Z$ along the helix axis. $E_n$ is the energy of the *n-th* molecular conformation; it is determined entirely from the intramolecular interactions within the solute. $V_n(\omega, Z)$ is the position-orientation-conformation dependent potential of mean torque (PMT); it is determined from the intermolecular interactions between the solute molecule and the molecules of the given domain of the solvent phase. The summation in eq (2) extends over all the conformations (assumed discrete and labelled by the index $n$) that are accessible to the solute molecule. The angular integration extends over all possible orientations $\omega$ of a set of molecular axes $x, y, z$, rigidly attached to the solute molecule, relative to a phase-fixed macroscopic frame $X,Y,Z$. The positional integration extends over one pitch length of the periodic $Z$-modulation of the solvent phase. The probability $p_n$ of the flexible solute molecule to be found in conformation $n$ within a domain of given handedness in the $N_{PT}$ phase is given according to eq (2) by the expression:

$$p_n \equiv \frac{e^{-E_n/kT}}{\zeta h} \int_0^h dZ \int d\omega e^{-V_n(\omega,Z)} \quad . \tag{3}$$

If during the *NMR* time scale the solute molecule diffuses rapidly enough to visit domains of opposite handedness, then the expression in eq (2) should be extended to include averaging over such domains. As shown in [26], domains of opposite handedness differ only with respect to the signs of certain terms (see eqs (11) and (12) below) forming the potential of mean torque $V_n(\omega, Z)$.

An expression analogous to eq (1) relates the order parameters to the spectra of $^{13}$C NMR presented in section 2. The counterpart of expression (1) for the spectroscopic observables in the case of proton residual dipolar couplings is

$$D_{ij} = -K_{ij} \left\langle r_{ij}^{-3} P_2(\hat{H} \cdot \hat{e}_{ij}) \right\rangle \quad , \tag{4}$$

where $r_{ij}$ denotes the distance between the $i, j$ pair of nuclei and $\hat{e}_{ij}$ is the direction of the inter-nuclear vector; $K_{ij}$ is the dipolar coupling constant for the $ij$ pair. Unlike the case of rigid molecules, where the inter-nuclear distance $r_{ij}$ is constant, for flexible molecules $r_{ij}$ could be strongly conformation dependent for certain nuclear pairs and therefore have a strong modulating effect on the averaging indicated in the right-hand side of eq (4).

According to eq (2) the differences in the averaging in different solvent phases enter solely through the PMT $V_n(\omega, Z)$. As shown for rigid solute molecules[26,30], the PMT can be derived from the solvent –solute intermolecular interactions and the molecular organization within the solvent phase. Explicit forms were derived for rigid solute molecules of different symmetries dissolvent in the common uniaxial nematic ($N_U$), the chiral nematic (N*) and $N_X$ solvents. In the remainder of this section we extend the formulation of the PMT to flexible molecules.

Strictly, one could apply the formulation used for rigid solutes to each molecular conformation of the flexible solute. Usually, however, this is not practically feasible due to the large number of conformations involved and their lack of sufficient symmetry elements. On the other hand, flexible molecules consisting of identical rigid repeat units can be more efficiently treated by approximate schemes which built the interaction of the entire molecule in terms of the interactions of its repeat units. Here we focus our attention primarily to the so-called segment-wise additive approximation, which has been successfully used in the past to describe flexible solute ordering in conventional nematic solvents in the context of the chord model.[33,40] Systematic inclusion of non-additive contributions is straight-forward but will not be considered here. The reason for not including non-additive contributions is that the available experimental data are not sufficiently detailed to allow the quantitative evaluation of the effects of such contributions on the solute ordering.

A flexible solvent molecule is described as a collection of rigid submolecular segments jointed in a specific fixed sequence and allowed to rotate relative to one another. A conformation, defined by the set of rotation angles (continuous or discrete) of each such segment relative to its neighbours (or relative to a fixed molecular frame) defines a molecular geometrical configuration of the jointed rigid segments and is assigned a particular energy value $E_n$, the conformational energy. Each rigid submolecular segment $s = 1, 2, \cdots$ is assigned a segmental axes frame $x_s, y_s, z_s$ and a PMT $V^{(s)}(\omega_s, Z_s)$ according to the formulations in refs [26,30] for rigid molecules. Here $\omega_s$ denotes the orientation of the segmental axes frame relative to the local phase directions $\hat{n}_h$ (the helical twist axis), $\hat{m}$ (the twisting polar director), $\hat{l}_h$ (perpendicular to the previous two). $Z_s$ indicates the position of the segment along the $Z$ axis of the helix. The orientation of the segment axes relative to the domain-fixed macroscopic frame $X,Y,Z$ is obtained by expressing the coordinates of the local phase directions in that frame according to [26,30]

$$\begin{aligned}\hat{m}(Z_s) &= -\hat{X} \sin k(Z_s + Z_0) + \hat{Y} \cos k(Z_s + Z_0) \\ \hat{l}_h(Z_s) &= \hat{X} \cos k(Z_s + Z_0) + \hat{Y} \sin k(Z_s + Z_0)\end{aligned} \quad . \tag{5}$$

Here $k$ is the wavenumber of the helical twisting and $Z_0$ is a constant phase within each domain.

Depending on the symmetry of the segment, we have for $V^{(s)}(\omega_s, Z_s)$ the three leading contributions[26,30]

$$V^{(s)}(\omega_s, Z_s) = V^{(s;1)}(\omega_s, Z_s) + V^{(s;2)}(\omega_s, Z_s) + V^{(s)*}(\omega_s, Z_s) \quad , \tag{6}$$

with

$$V^{(s;1)}(\omega_s, Z_s) = \left\langle g_{a_s}^{(1)} \right\rangle' (\hat{a}_s \cdot \hat{m}) \quad , \tag{7}$$

representing the first rank (polar) term and

$$V^{(s;2)}(\omega_s, Z_s) = \left\langle g^{(2)}_{a_s b_s} \right\rangle' S^{n_h n_h}_{a_s b_s} + \left\langle \Delta^{(2)}_{a_s b_s} \right\rangle' \left( S^{l_h l_h}_{a_s b_s} - S^{mm}_{a_s b_s} \right) + \left\langle \Phi^{(2)}_{a_s b_s} \right\rangle' \left( S^{n_h l_h}_{a_s b_s} + S^{l_h n_h}_{a_s b_s} \right) \quad , \quad (8)$$

representing the second rank achiral term. The chiral (pseudovector) term $V^{(s)*}(\omega_s, Z_s)$ appears in principle due to chiral imbalance induced to the (statistically achiral) dimer molecules by the twist modulation of the polar ordering in the $N_{PT}$ phase. This chiral imbalance would obviously be absent in case of $N_{PT}$ phases formed by strictly rigid molecules [41] and is estimated to be small in the case of flexible molecules. [24,25] Accordingly, the chiral term in eq (6) has been argued [26] to have a marginal quantitative effect on the solute ordering and will be omitted in the calculations to follow. The tensor quantities appearing in the expressions of eq (8) are defined as $S^{IJ}_{a_s b_s} \equiv \frac{3}{2}(\hat{a}_s \cdot \hat{I})(\hat{b}_s \cdot \hat{J}) - \frac{1}{2}\delta_{IJ}\delta_{a_s b_s}$, with $I, J$ standing collectively for any of the local $N_{PT}$ phase axes $\hat{n}_h, \hat{l}_h, \hat{m}$, and $a_s, b_s$ denoting collectively any of the segmental frame axes $x_s, y_s, z_s$. To express the polar and apolar segmental contributions of eqs (7), (8) in a common macroscopic frame, according to eq (5), we introduce a molecular frame $x,y,z$ fixed on some conveniently chosen segment of the molecule and denote by $\vec{R}$ the position of its origin in the macroscopic frame $X,Y,Z$. Then, with $\vec{R} \cdot \hat{n}_h = Z$ and with the intramolecular vector $\vec{r}_s$ defining the position of the origin of the frame of segmental axes $x_s, y_s, z_s$ in the molecular frame $x,y,z$, we have for the coordinate of the $s^{th}$ segment along the helix axis $Z_s = Z + \vec{r}_s \cdot \hat{n}_h$. It is noted from this expression that the explicit segmental, orientational and conformational dependence of $Z_s$ is introduced through the projection of intramolecular vector $\vec{r}_s$ on the helix axis $\hat{n}_h$.

In the additive approximation, the PMT for the entire solute molecule, appearing in eq (2), is simply the sum of the contributions from all the rigid submolecular segments forming the flexible molecule:

$$V_n(\omega, Z) = \sum_s V^{(s)}(\omega_s, Z_s) \quad . \quad (9)$$

The segments in all the solvent and solute molecules considered in this study can be decomposed into just two kinds of submolecular segments: (i) $CH_2$ segments, either as parts of the *n*-decane solute, or as parts of the spacer of the mesogenic dimers and (ii) CB mesogenic units, with their linkage to the spacer chain of the CB-dimers. The first and second rank contributions pertaining to these segments, according to eqs (6), (7), as used in the present modeling of the additive PMT, are presented below.

**A. Alkane segments.** Each $CH_2$ segment is represented by the two half-bonds of the carbon atom with the adjacent carbons (C-C-C "isosceles triangle") and by the two CH bonds (see figure 4). The base of the triangle is chosen to be the $y_s$ axis of the segment and the height of the triangle is assigned as the $z_s$ (polar) axis; the $x_s$ axis is then normal to the plane of the triangle (i.e., parallel to the base of the triangle formed by the H-C-H bonds). Such a segment then has a C$_2$ axis, the $z_s$ (polar) axis, and two planes of symmetry ($z_s - x_s, z_s - y_s$) containing that axis. This means that the segmental interactions are invariant with respect to the independent inversions $x_s \Leftrightarrow -x_s ; y_s \Leftrightarrow -y_s$.

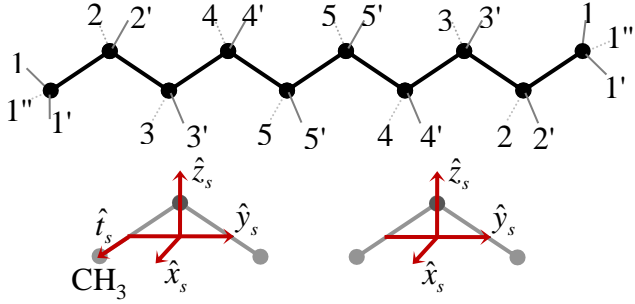

**Figure 4.** (top) Atomic diagram of the n-decane molecule in the all trans conformation, showing the labelling of the hydrogens of the methyl and methylene groups. (bottom) Geometry of the terminal CH$_3$-CH$_2$-CH$_2$ (left) and the inner CH$_2$-CH$_2$-CH$_2$ (right) segments and local axes used in the formulation of the segmental PMT.

Then, according to eqs (7, 8) we have for the polar part:

$$V^{(1)s}_{chord} = p_{CH_2}(\hat{z}_s \cdot \hat{m}) , \qquad (10)$$

and for the uniaxial quadrupolar part:

$$V^{(2)s}_{chord} = \alpha S^{n_h n_h}_{y_s y_s} + \beta \left( S^{l_h l_h}_{y_s y_s} - S^{mm}_{y_s y_s} \right) + \gamma S^{n_h l_h}_{y_s y_s} . \qquad (11)$$

The parameter $\gamma$ changes sign with the handedness of the twisted domain [26] whereas the other parameters, $p_{CH_2}, \alpha, \beta$ are invariant to handedness inversion.

The n-decane molecule is terminated by two CH$_3$ groups. These are treated as polar and uniaxial, with respective contributions $V^{(1)s}_{term} = p_{CH_3}(\hat{t}_s \cdot \hat{m})$ and $V^{(2)s}_{term} = \alpha_t S^{n_h n_h}_{t_s t_s} + \beta_t \left( S^{l_h l_h}_{t_s t_s} - S^{mm}_{t_s t_s} \right) + \gamma_t S^{n_h l_h}_{t_s t_s}$, with fixed (temperature independent) coefficient ratios $p_{CH_3}/p_{CH_2}, \alpha_t/\alpha, \beta_t/\beta, \gamma_t/\gamma$. Specifically [33], $\alpha_t$ is related to $\alpha$ through the $CCC$ bond angle according to $\alpha_t = \dfrac{\alpha}{2(1-\cos(CCC))}$.

In a uniaxial apolar nematic phase one obviously has vanishing values for the parameters $p_{CH_2}, p_{CH_3}, \beta, \beta_t, \gamma, \gamma_t$. In this case the PMT reduces to that of the chord model [32,40] in its simplest (additive) form.

**B. CB mesogenic units:** These are assumed to consist of a uniaxial and longitudinally polar CB cores, represented by a unit vector **L,** rigidly attached to which are linkage chords (see figure5) interacting with the solvent medium as in **A** above. The potential of mean torque associated with such uniaxial CB cores is of the form:

$$V_{C-B} = p_{C-B}\left(\hat{L}\cdot\hat{m}\right) + A S_{LL}^{n_h n_h} + B\left(S_{LL}^{l_h l_h} - S_{LL}^{mm}\right) + C\left(S_{LL}^{n_h l_h} + S_{LL}^{l_h n_h}\right) \quad (12)$$

Inversion of the handedness of the domain twisting implies the change of sign of the *C* parameter in eq (12), the other parameters being invariant to handedness inversion. In a uniaxial apolar nematic medium, all the terms in eq (12) vanish except for the second rank term proportional to *A*.

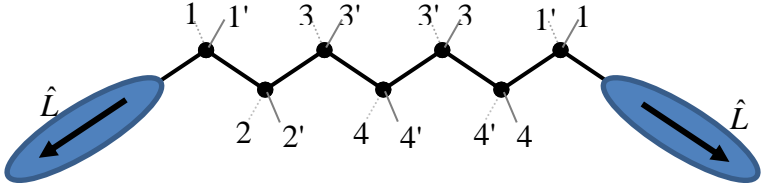

**Figure 5.** Geometry of CB7CB dimers showing the labelling of the axes of the CB submolecular units used in the formulation of the PMT and the numbering of the spacer hydrogens used in the text.

In the case of mesogenic dimers, non-additive PMT terms formed by the two mesogenic units are significant, although neglected in the present calculation. In particular, a polar term accounting for the steric dipole formed by the two mesogenic units in non-linear configurations is suggested by the molecular model in ref [26]. Such a term results from the polar correlations of a pair of mesogenic units attached at the two ends of the spacer chain of a single dimer. Denoting these mesogenic units by the unit vectors $\hat{L}^+, \hat{L}^-$ and the direction of vector jointing their midpoints by $\hat{d}$, this term has the form:

$$V_{C-Bpair} \sim \left(\hat{d}\cdot\hat{n}_h\right)\left[\left(\hat{L}^+\cdot\hat{n}_h\right)\left(\hat{L}^+\cdot\hat{m}(Z^+)\right) - \left(\hat{L}^-\cdot\hat{n}_h\right)\left(\hat{L}^-\cdot\hat{m}(Z^-)\right)\right] \quad . \quad (13)$$

Here the possibility of the two mesogenic units being found at solute locations $(Z^+, Z^-)$ corresponding to different directions of the twisting of the polar axis $\hat{m}$, is explicitly indicated. Obviously, when the two mesogenic units are not collinear, the above term confers to the PMT a polar aspect even if these units are strictly apolar.

## 4. Determination of the order parameters and comparison with NMR measurements.

Using the PMT parts as formulated in eqs (10)-(11) we compute the averages indicated in eqs(1) and (4) according to the averaging procedure of eq(2). These are then compared to the respective quantities obtained from the NMR measurements that are available in the N and $N_X$ phase of the CB7CB and CB9CB dimers. These measurements comprise decane quadrupolar splittings presented in section 2 and residual dipolar couplings of the CB7CB dimer in its neat nematic phases obtained from [35].

There are also available NMR data, restricted to quadrupolar splittings of the $\alpha,\omega$ positions of the CB7CB spacer[2], quadrupolar splittings of 8CB, dissolved in the same dimer phases, and deuteriated at the $\alpha$-position of the pendant chain [2] and also for the asymmetric dimer CB6OCB. [42] . Whilst these are straightforward to analyse in the same framework, they are not considered in this work because the information they convey is rather limited and their theoretical reproduction poses no challenge.

First, we note that, in all the cases to be analysed, the spectra are obtained for magnetically aligned samples, i.e. with the helix axis $\hat{n}_h$ of the $N_X$ phase, and hence with the macroscopic Z axis, held parallel to the magnetic field. Accordingly, the quantities to be averaged in eqs(1) and (3) involve only the Z-projections of the molecular site vectors $\hat{e}_s$ and are therefore unaffected by the Z-dependent rotation of the $\hat{m}$, $\hat{l}_h$ local axes (see eq 5). Second, the above set of available experimental data from NMR is not detailed enough to provide a sufficiently restrictive testing of the parameterization used for the PMT parts in eqs(9-11). We therefore proceed with simplifying approximations to reduce the number of independent parameters of the PMT, and test the transferability of these parameters among the different solute-solvent systems and temperatures. As the polar interactions are an essential feature in the $N_X$ phase, no significant simplification can be applied to the polar contributions. The non-polar contributions, on the other hand, are susceptible to further simplification in the context of a crude approximation. To this end we consider the second rank part of the PMT for the uniaxial core segment in eq(11):

$$V_{CB}^{(2)} = A S_{LL}^{n_h n_h} + B\left(S_{LL}^{l_h l_h} - S_{LL}^{mm}\right) + C\left(S_{LL}^{n_h l_h} + S_{LL}^{l_h n_h}\right) , \qquad (14)$$

and assume that the interaction tensor $S_{LL}^{IJ}$ is uniaxial in its principal axis frame. This is to some extent justified by the small values obtained for the biaxiality in an explicit molecular model of primitive dimesogens leading to $N_{PT}$ phase ordering.[24] Since $\hat{m}$ is the local director of the phase, the principal axis frame is obtained by rotating $\hat{n}_h, \hat{l}_h$ about $\hat{m}$. Denoting this rotation angle by $u_L$, the condition of uniaxiality relates at any given temperature the parameters B and C to the parameter A through the angle $u_L$ according to

$$B = A\frac{(1-\cos 2u_L)}{1+3\cos 2u_L}; \quad C = -2A\frac{\sin 2u_L}{1+3\cos 2u_L} \ . \tag{15}$$

Thus the three parameters are reduced to two, which can be conveniently chosen to be $A$ and $u_L$. In accordance with the behaviour of the $A, B, C$ parameters discussed in section 3, it is clear that inversion of the handedness of the domain twisting sense implies the change of sign of the $u_L$ angle.

It is further assumed, for simplicity, that the second rank tensor of the chord segment potential in eq(11) is also uniaxial in its principal axis frame. The rotation angle about $\hat{m}$ by which the chord interaction tensor is diagonalized is denoted by $u_c$. The parameters $\alpha, \beta$ and $\gamma$, of eq(11) are related among themselves analogously to eq(15). Lastly on assuming a constant (temperature independent) ratio for the magnitude $\alpha/A$, the entire second rank part of the PMT is specified by just two temperature-dependent parameters, $\alpha$ and $u_c$, for the alkane solutes. For the dimer solutes, this set is augmented by $u_L$ and a constant ratio $\alpha/A$. Clearly these are only approximations, aimed at handling the limited information presently available from measurement, and it is by no means implied that they are characteristics of the $N_X$ phase. No such approximations where required, nor made, aside from the neglect of the marginal biaxiality, in the case of rigid solutes in the $N_X$ phase. Accordingly, these simplifications should not cause any confusion with the $N_{TB}$ model wherein the presence of a unique local axis, which is a common principal axis of all the tensor quantities, is an exact property of the model: in the $N_{TB}$ case the common principal axis is the nematic director $\hat{n}$, undergoing the heliconical deformation, and due precisely to the full rotational symmetry about the local $\hat{n}$ all the molecular tensor quantities are uniaxial and there can be no transversely polar interaction of the solute molecules with the solvent medium.

Turning now to the polar terms, a simplifying approximation, in the spirit of the constant $\alpha/A$ ratio, is to treat the ratio of $p_{chord}/p_{core}$ as a temperature independent constant. For the same reasons of simplification, the pitch $h$ (equivalently, the wave vector $k=2\pi/h$) of the helical twisting of the polar director $\hat{m}$, which determines the local environments sensed by the different segments of the extended solutes, is taken to be fixed, at some average value obtained from experiments, [3,13,43] despite the well-known, albeit limited temperature dependence of this quantity ($h \sim$ 8-12 nm). Lastly, the conventional 3-state RIS model is used for the generation of the conformations of alkyl chains and spacers, with standard [33] values for the conformational energy parameters ($E_{tg} = 0.9$ kcal/mol), the torsion angles ($\varphi_g = 114°$), bond lengths and bond angles ( ($HCH = 110°$ ; $CCC = 113°$).

*4.1 Quadrupolar NMR splittings of decane in CB7CB and CB9CB*

Figures 6 and 7 show the results of the calculation of the splittings of all the deuterated sites of decane as a solute in the nematic phases of CB9CB and CB7CB as functions of temperature. The splittings in the low teperature $N_X$ phase are calculated according to the $N_{PT}$ model and assuming that the twisted domains are large enough so that a decane solute molecule resides within a domain of given handedness during the entire NMR measurement time scale. The optimal values of the fitting parameters $p_{CH_2}$, $\alpha$ and $u_c$ are plotted in figure 6(b) and 7(b). The agreement with the measured values is seen to be complete, within experimental accuracy.

The respective values of the splittings in the high teperature N phase are calculated in two ways:

(i) Using the $N_{PT}$ model, as in the case of the $N_X$ phase, with the only difference that the solute molecule is now assumed to difuse to multiple domains of opposite handednes within the NMR time window. Accordingly the measured splitting is obtained as the average of the slittings that woud be obtained for each domain, thus eliminating any enantiotopic discrimination from the spectra.
(ii) Using the uniaxial, apolar, achiral nematic ($N_U$) parmeterisation of the PMT, according to which there is no polar contribution ($p_{CH_2}=0$) and no "tilt" angle ($u_c$); hence no enantiotopic discrimination in the spectra.

The results of the optimal calculated splittings in the two cases and the comparison with the expreimental values, presented in figure 6(a), show that a clearly better agreement (practically complete) is obtained with the model in (i), while the results with model (ii) show systematic deviations for $CD_1$ and $CD_2$. The optimal values of the fitting parameters in the high teperature N phase for each model are shown in figure 6(b). There it is apparent that the variation of the fiting parameters for the model in (i) is continuous across the N-$N_X$ phase transition whilst for the model in (ii) there are discontinuities in all three fitting parameters.

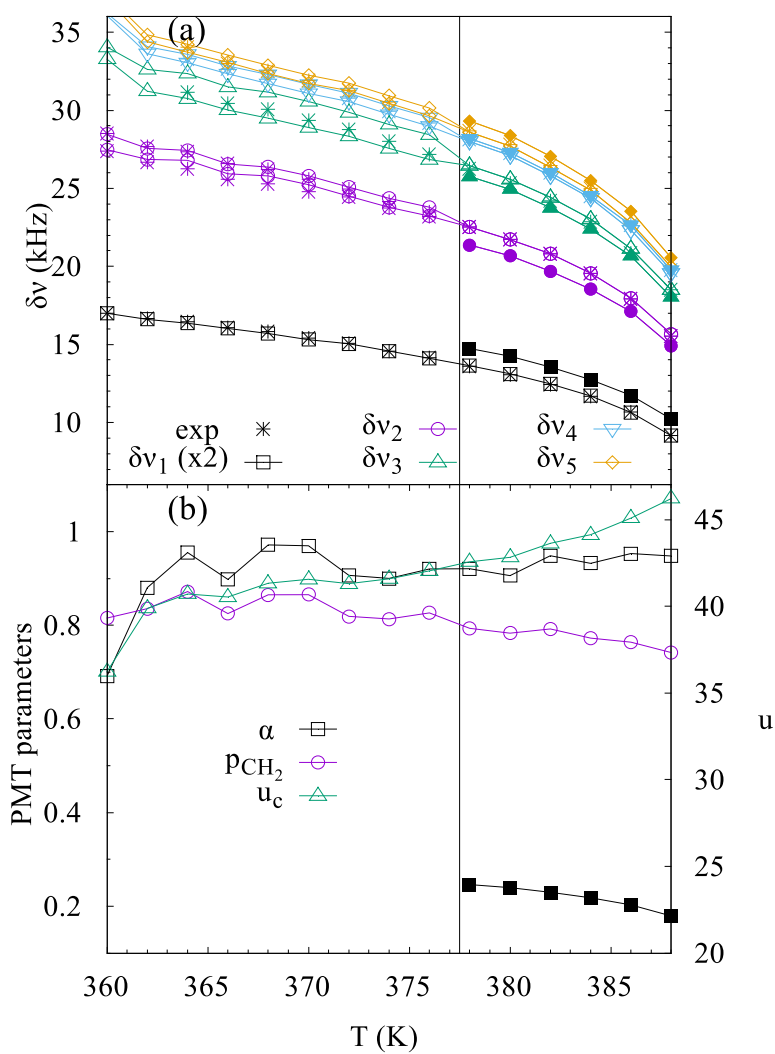

**Figure 6.** (a) Calculated and experimentally measured splittings of perdeuteriated decane (decane-d22 in the two nematic phases of the dimer CB9CB. The (*) symbols represent the measured quadrupolar splittings. Open symbols represent the calculated splittings. The Nx-N transition temperature is marked by the vertical line. The splittings in the Nx phase are calculated according to the $N_{PT}$ model. The splittings in the high temperature N phase are calculated according to (i) the $N_{PT}$ model (filled symbols) and (ii) the uniaxial, apolar, achiral nematic model (filled symbols) as detailed in the text. (b) Optimal values of the fitting parameters of the calculated spectra In the $N_X$ phase and the in the high temperature N phase (open symbols for the calculations according to model (i) and filled symbols for model (ii) above).

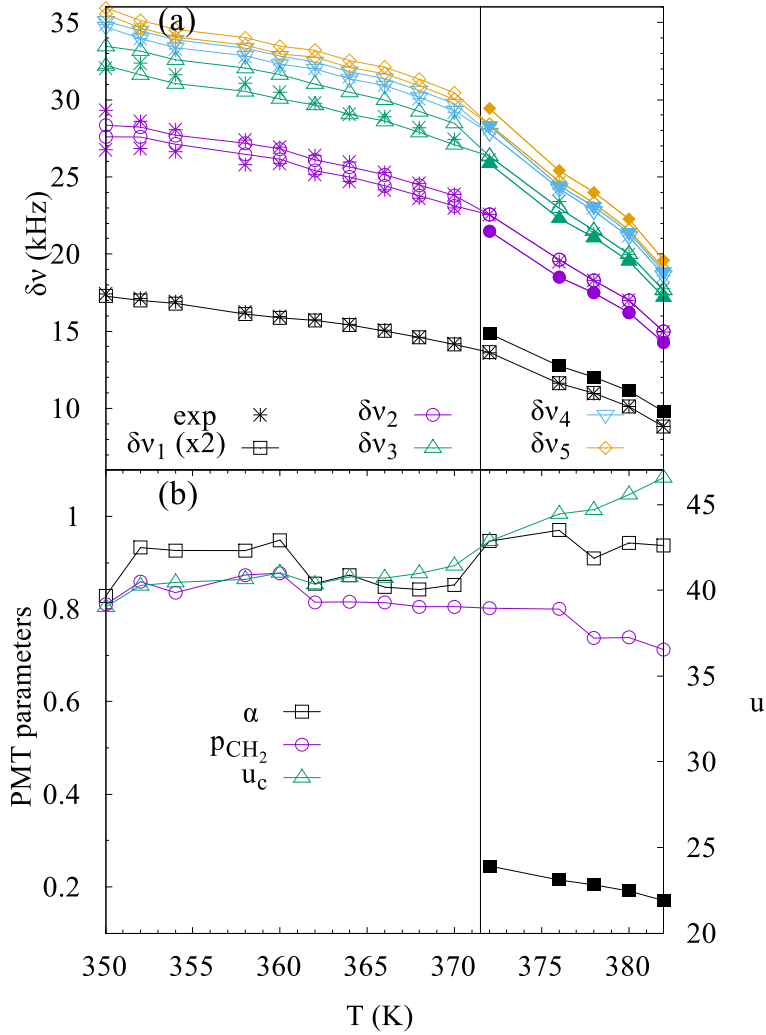

**Figure 7.** as in Figure 6 but for the dimer CB7CB.

Using the PMT parameter values of figure 4.1, it is possible to evaluate, according to eq (3), the probabilities of all the conformations of the decane solute in the nematic phases of CB9CB. A large number of these conformations are chiral and come in pairs of opposite handedness; in an achiral medium the opposite handedness conformations have equal probabilities. However, in a chiral environment this equality of probabilities does not hold. Thus, within a twisted $N_{PT}$ domain, a pair of opposite handedness conformations, denoted by $n_+, n_-$, will generally have different probabilities $p_{n_+} \neq p_{n_-}$. A measure of the overall probability imbalance induced by the chiral medium is given by the parameter $\Delta^* = \sum_n \left| p_{n_+} - p_{n_-} \right|$, which varies between 0 (for no chiral imbalance) and 1 (for complete elimination of all conformers of a given handedness). The calculated values of $\Delta^*$ vary slightly with temperature and are found in the range $\Delta^* = 0.6 \times 10^{-3}$ to $0.8 \times 10^{-3}$ in the $N_X$ phase and $\Delta^* = 0.8 \times 10^{-3}$ to $1.0 \times 10^{-3}$ for the $N_{PT}$ cluster picture of the N phase. This indicates that the twisted environment induces a very small chiral imbalance to the decane solute.

## 4.2 Calculation of the residual dipolar couplings of the CB7CB dimer in its neat nematic phases.

A detailed set of the residual dipolar couplings (RDC) between carbon and hydrogen nuclei has been obtained in refs [2,35] using proton-encoded $^{13}C$ 2D NMR experiments. The RDCs are measured at two temperatures corresponding to the high- and low-temperature nematic phases, N and $N_X$ respectively. These data offer the most detailed and direct description of the segmental ordering of the spacer of the CB7CB dimer, albeit at only two temperatures.

The theoretical reproduction of these measurements in the $N_X$ phase is performed by keeping the parameter ratios fixed to the temperature independent values $\alpha / A = 0.25$ and $p_{chord} / p_{core} = 0.6$. The optimal parameters of the PMT obtained from the fitting of the RDCs of the $N_X$ phase (at the temperature of 84°C) are $A = 2.635$, $p_{core} = 0.89$, $u_L = 25.7°$, $u_c = 40.2°$. The calculated RDCs are presented in Table I. With this parametrization the calculated order parameter of mesogenic units (with respect to the magnetic field direction) is $S = 0.32$, which compares fairly well with the experimentally determined, see figure 2, order parameter of the CN group at the same temperature $S_{CN} = 0.34$.

**Table I.** Experimental Residual Dipolar Couplings, $D_{i,j}/Hz$, obtained in [35] for CB7CB in the Nx phase at 84°C and their optimal calculated values.

|  | $(CH)_1$ | $(CH)'_1$ | $(CH)_2$ | $(CH)'_2$ | $(CH)_3$ | $(CH)'_3$ | $(CH)_4$ |
|---|---|---|---|---|---|---|---|
| exp[a] | 4881 | 3655 | 4145 | 3777 | 3588 | 2879 | 3247 |
| exp[b] | 4881 | 3655 | 3588 | 2879 | 4145 | 3777 | 3247 |
| Calc | 4864 | 3676 | 3557 | 2878 | 4208 | 3707 | 3261 |
| Err (%) | 0.4 | 0.6 | 0.9 | 0.1 | 1.5 | 1.9 | 0.4 |
| [a] Dipolar couplings with the original assignment in [35] where it is stated that the assignments in positions 2 and 3 could be reversed. ||||||||
| [b] Assignment of the measured residual dipolar couplings according to the optimal fitting. ||||||||

In the high temperature nematic phase both models of the phase, the $N_U$ and the cluster $N_X$ model, give satisfactory fittings. These are shown in table II. In the $N_U$ model, the optimal fitting, with $A$ and $\alpha$ allowed to vary independently, is obtained for $A = 1.47$ and $\alpha = 0.235$, giving less than 2% error for the RDCs. The calculated order parameter is $S = 0.3$.

In the cluster model, the optimal fitting (keeping fixed ratios $\alpha/A = 0.25$, $p_{chord}/p_{core} = 0.6$) is obtained for $A = 1.582$, $p_{core} = 0.40$, $u_L = 14.5°$, $u_c = 34.1°$. The calculated order parameter value is $S = 0.3$.

**Table II.** Experimental Residual Dipolar Couplings, $D_{i,j}/Hz$, obtained in [35] for CB7CB in the high temperature N phase at 106°C and their optimal calculated values.

|  | $(CH)_1$ | $(CH)_2$ | $(CH)_3$ | $(CH)_4$ |
|---|---|---|---|---|
| Exp | 3735 | 2339 | 3034 | 2264 |
| Calc (i) | 3719 | 2385 | 3036 | 2235 |
| Calc (ii) | 3722 | 2384 | 3028 | 2239 |
| (i) $N_U$ nematic model | | | | |
| (ii) Cluster $N_X$ nematic model | | | | |

Evaluation of the chiral imbalance through the $\Delta^*$ parameter defined earlier, using the optimal values determined from the fits in tables I and II, gives $\Delta^* = 0.31$ for the fits corresponding to table I and $\Delta^* = 0.12$ for those of table II with the cluster $N_X$ model. These values indicate that the tendency towards chiral discrimination of enantiomeric molecular conformations of the dimer molecule are appreciable and therefore that the pseudovector contributions to the PMT in eq (6) are not entirely negligible. It should be kept in mind, however, that these fits are obtained after the drastic simplifications dictated by the need to reduce the number of the adjustable parameters in view of the limited data available for fitting. In particular, the omission of non-additive terms formed by the two mesogenic cores of the dimer molecule, such as the term in eq (13), could influence significantly the calculations of quantities like $\Delta^*$ without dramatically influencing the RDCs of the spacer.

Lastly, a comment about the low values of the core order parameters obtained both from experiment and calculation: These values are very low compared to those found in the nematic phase of conventional calamitic mesogens. However, a meaningful comparison should take into account that, while the maximum value of the order parameter for such molecules is 1 (corresponding to perfect alignment of the molecules along a given direction), the maximum value for a V-shaped molecule is significantly less than 1 as perfect alignment of both mesogenic "wings" along a given direction is geometrically impossible. An accordingly normalized measure of the order parameters of the mesogenic cores is of the dimers is provided by the reduced order parameters

$$S_{RED} = \frac{1}{2} \left\langle \frac{P_2(\hat{Z} \cdot \hat{L}^+) + P_2(\hat{Z} \cdot \hat{L}^-)}{s_{max}(\hat{L}^+, \hat{L}^-)} \right\rangle \quad , \tag{16}$$

where $s_{max}(\hat{L}^+, \hat{L}^-)$ is the maximal orientational order of a molecular conformation with its mesogenic cores fixed at directions $\hat{L}^+, \hat{L}^-$. The conformation-depended quantity $s_{max}$ is a rotational invariant and is calculated as the maximum eigenvalue of the ordering tensor $Q_{ab} = (3L_a^+ L_b^+ - 1)/2 + (3L_a^- L_b^- - 1)/2$.

Calculation of the reduced order parameters using eq (16) gives the value $S_{RED} = 0.56$ for the fits in table I and $S_{RED} = 0.50$ for those of table II with the $N_X$ cluster model and $S_{RED} = 0.49$ for the $N_U$ model. These values are in the usual range found for conventional calamitic compounds and also show the normally observed trend with temperature, i.e. increasing with decreasing temperature.

**5. Discussion and Conclusions.**

The experimental NMR information presented in section 2, comprising the probing of segmental ordering of the flexible and extended solute decane-$d_{22}$ in the nematic phases formed by a range of mesogenic CBnCB dimers as solvents, as well as the measurement of the ordering of the mesogenic units of these dimers, offer new insights into the low temperature Nx phase exhibited by odd-spacer length dimers. These observations expand significantly the perceptions gained from previous NMR studies using small rigid solutes[2,31] and selectively deuteriated nCB solutes[2], as well as apparent insights from NMR measurements addressing directly the segmental ordering of the spacer in CBnCB molecules. [2,4] The fact that this broad collection of experimental findings is rationalized and reproduced coherently by the polar-twisted nematic ($N_{PT}$) shows that this model correctly describes the local symmetries and the molecular ordering, and short-pitch modulations thereof, in the $N_X$ phase. Viewed separately, some of the above sets of measurements, not being sufficiently detailed and hence necessarily restrictive, could be described and reproduced in different ways.[35,44] The simultaneous reproduction, however, of all the above measurements, in terms of a small number of $N_{PT}$ model-parameters with high degree of transferability, as demonstrated in section 4, suggests that a physically-sound and consistent description of the $N_X$ phase has been achieved. This conclusion might prompt further, more detailed, experimental studies of segmental orientational order in the $N_X$. Such studies could, on one hand, test the $N_{PT}$ model more severely. On the other hand, they could provide the opportunity for the deployment of the full parameterization of the model. As detailed in section 4, a simplification of this parameterisation was dictated by the relatively low information content of even the entire collection of the presently available experimental NMR studies.

Additionally, the successful description of the molecular ordering in the $N_X$ phase in this study enabled a better understanding of the nature of the high temperature N phase, that in turn, lead to the suggestion that the latter N phase consists of clusters of the same structure as the polar, twisted $N_X$ domains, albeit of much smaller spatial extent.. Aside from the successful reproduction of the quadrupolar splittings and the RDCs, the assignment of the high temperature nematic phase (N) as a cybotactic cluster polar twisted phase ($N_{CybPT}$) is consistent with and/or supported by a number of independent observations listed below:

(i) The appearance of a maximum in the order parameter of the mesogenic units (Fig.2) followed by a decreasing tendency with decreasing temperature. This observation, together with the lack of enantiotopic discrimination in the N phase has been alternatively interpreted in a variety of ways. For example, picturing the N phase as a) a "tilted" nematic (similarly to the $N_X$, in order to associate the non-monotonic temperature dependence of the ordering with a variation of the "tilt") but not "chiral" (unlike the $N_X$, in order to justify the lack of enantiotopic discrimination); b) as a splay-bend phase [45]; or c) a phase consisting generally of non-chiral molecular aggregates. [46]

(ii) The qualitative similarity (and marginal quantitative difference) of the X-ray diffractograms across the N-$N_X$ phase transition[2,12,39], indicating no change in the local (i.e. nanometer scale) molecular organization.

(iii) The practically continuous variation of the measured dielectric anisotropy [10] across the N-$N_X$ phase transition, also indicating invariance of the local structure.

(iv) The strong effect of an applied magnetic field on the I-N transition, [47] which can be interpreted as a result of the field addressing structured clusters rather than individual molecules. [48–52]

(v) Molecular simulations [25] reproducing the essential features of the $N_X$ phase and showing no changes in the local position-orientation molecular correlations at the N-$N_X$ phase transition.

(vi) The dramatic increase of the reorientational viscosity across the N-$N_X$ phase transition, allowing the measurement of NMR spectra with the director perpendicular to the magnetic field in the $N_X$ phase but not in the N phase. [4] This effect would be readily explicable in terms of the growth of polar-twisted clusters into macroscopic-size domains.

Although on its own, none of the above listed observations constitutes a definitive proof of the N = $N_{CybPT}$ assignment, the combination constitutes a very strong indication that this is a cybotactic phase.

The cluster character of the high temperature N = $N_{CybPT}$ would naturally imply profound differences in the treatment of the deformation elasticity compared to the treatment of common ("molecular" uniaxial, apolar) nematics in the context of the Frank-Oseen elasticity theory. In particular, the often reported decrease of the bend elastic constant $K_3$ on approaching the $N_X$ phase, typically measured through the threshold voltage in the Fredericks effect[3] would simply reflect the ease of realignment of clusters as the latter grow larger.

Lastly, the possible $N_{PT}$-cluster character of the high temperature nematic phase of odd-dimers generally might entail a radically different rationalization of the well-known enhanced even-odd effect in the variation of the Nematic-to-Isotropic transition temperature for dimers. [53–55] In this case the dramatic variation could be related to the transition from isotropic to structurally different nematic phases, depending on the parity of the spacer: Conventional uniaxial nematic ($N_U$) for even spacers vs $N_{CybPT}$ for odd spacers.

**Appendix A.**

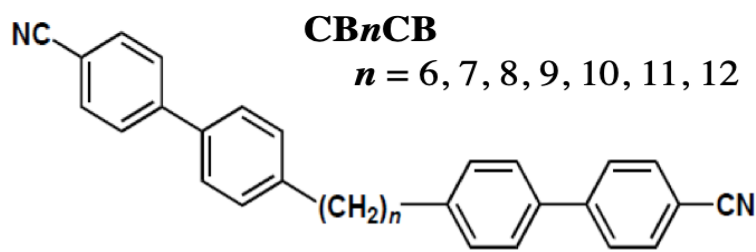

| Dimer  | n  | Phase Transitions          | Phase Type |
|--------|----|----------------------------|------------|
| CB6CB  | 6  | Cr 191 N 231 I             | N          |
| CB7CB  | 7  | Cr 102 $N_x$ 104 N 116 I   | N, $N_X$   |
| CB8CB  | 8  | Cr 178 N 199 I             | N          |
| CB9CB  | 9  | Cr 85 $N_x$ 108 N 124 I    | N, $N_X$   |
| CB10CB | 10 | Cr 140 N 172 I             | N          |
| CB11CB | 11 | Cr 105 ($N_x$ 84) N 125 I  | N, $N_X$   |
| CB12CB | 12 | Cr 139 N 159 I             | N          |

**Table A.1.** Chemical structure of CBnCB and respective transition temperatures taken from the first DSC heating scans (10 Kmin$^{-1}$; see ref [56]) A complete homologus series is shown with the shaded dimer names constituting the subjects of investigation in this work.

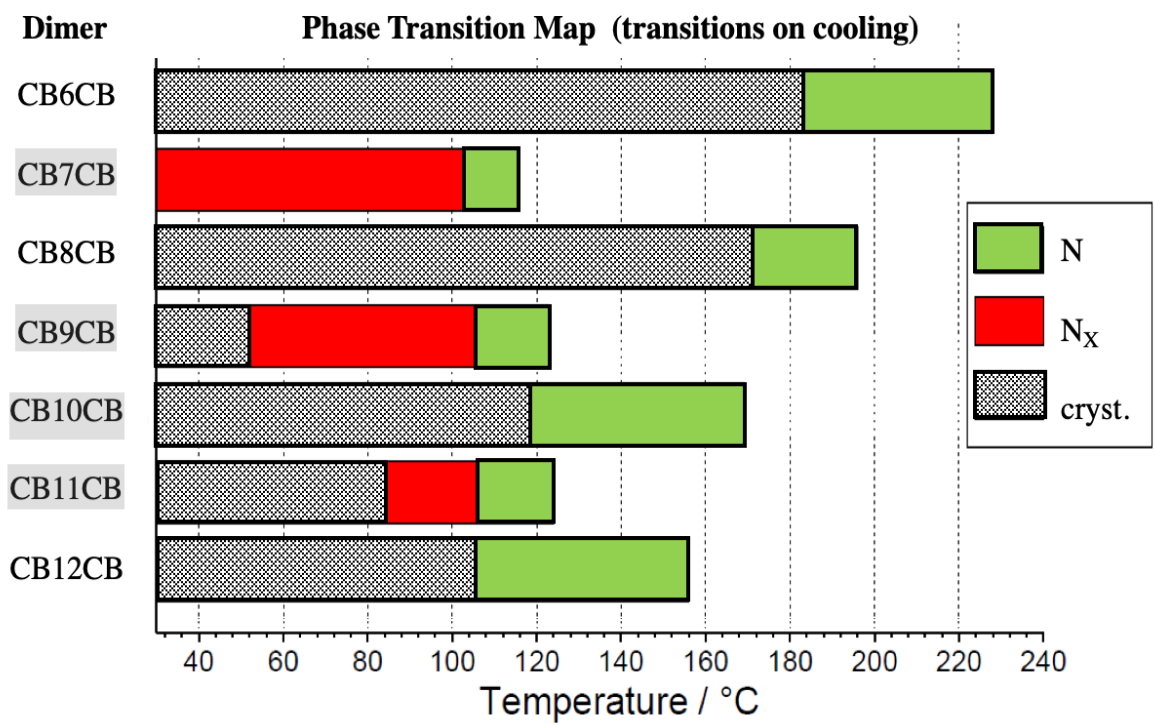

**Figure A.1** Transition temperatures of CBnCB dimers from first DSC cooling scans (10 Kmin$^{-1}$).

**Acknowledgements**.

Z.A. and C. W. acknowledge funding through the EPSRC projects EP/J004480 and EP/M015726 .

## Supplementary Information

Figure SI.1 shows the $^2$H NMR spectra of decane-$d_{22}$ in CB11CB for selected temperatures. In Figure SI.2, the quadrupole splitting of the methyl group and its adjacent methylene—the deuterons on the penultimate carbons of decane—are plotted against reduced temperature. The transitions from nematic to isotropic $T_{N-I}$, nematic to lower temperature nematic, $T_{N-Nx}$ and from Nx to crystal, $T_{Nx-K}$, (the vertical dashed lines on the reduced temperature axes) are computed from the data Tables at the conclusion of the SI. As the temperature decreases into the Nx phase, the slope of the temperature dependence of the splitting increases, but there is no doubling of the quadrupolar splitting.

Similar data is shown in Figures SI.3 - 6 for CB9CB and CB7CB respectively.

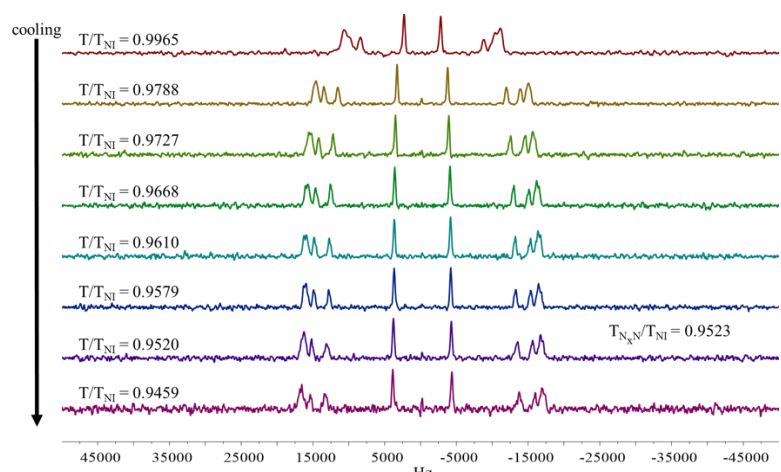

**Figure SI.1.** $^2$H NMR spectra of decane-$d_{22}$ in CB11CB at decreasing temperatures (from top to bottom). The $N_x$-N transition is noted and occurs at a reduced temperature of 0.9523.

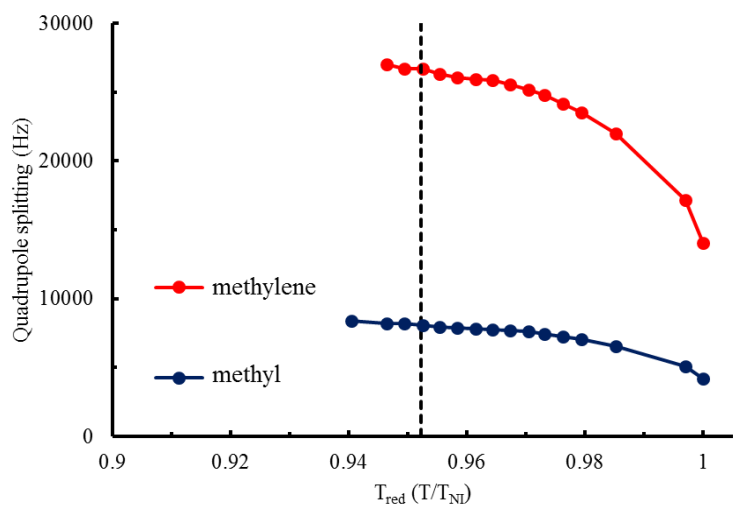

**Figure SI.2.** The temperature dependence (in reduced temperature) of the quadrupole splitting of the methyl and its adjacent methylene group of decane-d$_{22}$ in CB11CB with the N to Nx phase transition shown as dotted line.

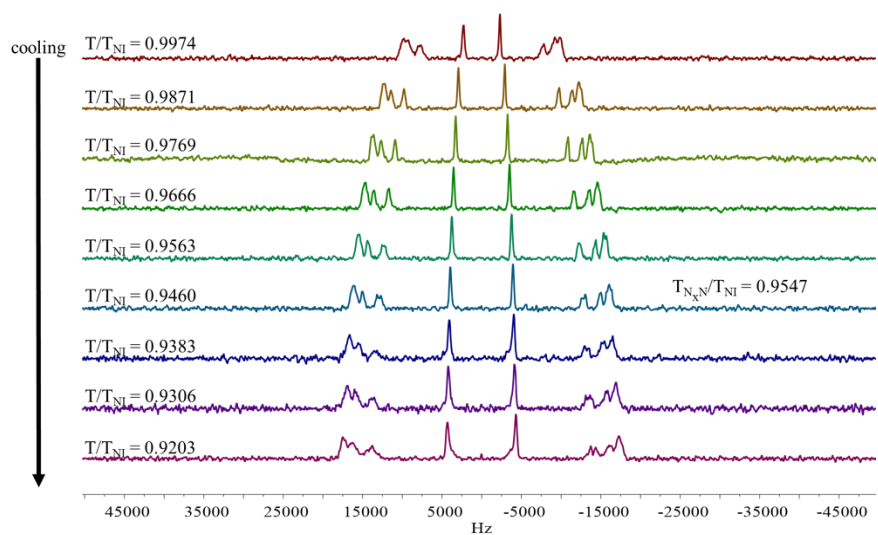

**Figure SI.3.** $^2$H NMR spectra of decane-d$_{22}$ in CB9CB at decreasing temperatures (from top to bottom) The N$_x$-N transition is noted and occurs at a reduced temperature of 0.9547.

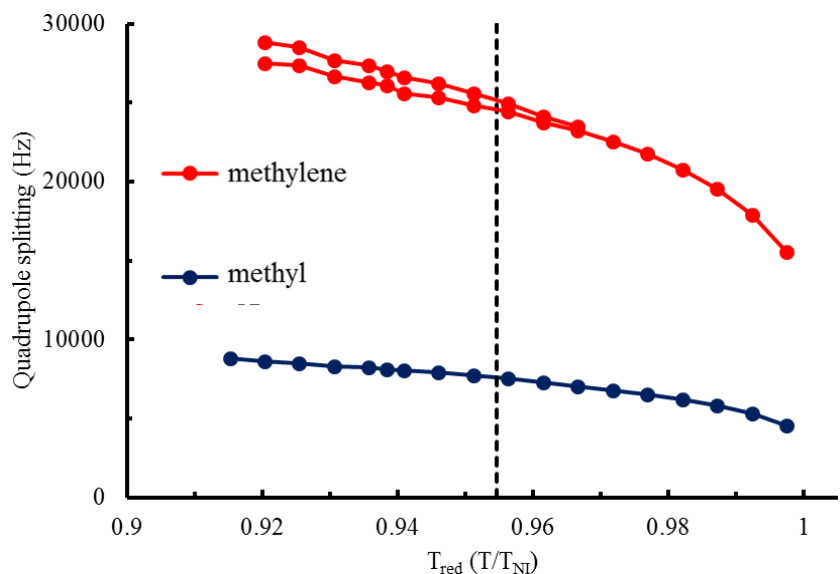

**Figure SI.4.** The temperature dependence (in reduced temperature) of the quadrupole splitting of the methyl and its adjacent methylene group of decane-$d_{22}$ in CB9CB with the N to Nx phase transition shown as dotted line.

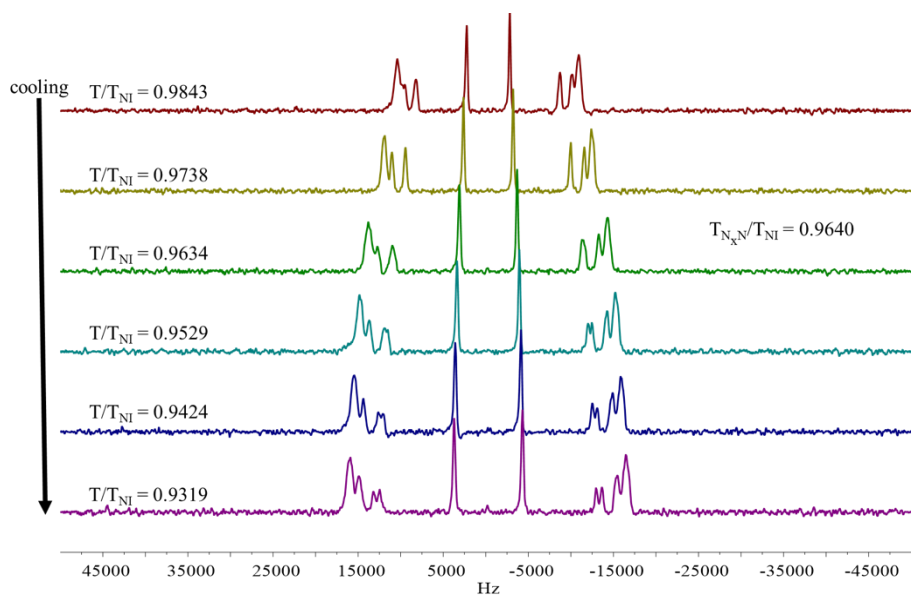

**Figure SI.5.** $^2$H NMR spectra of decane-$d_{22}$ in CB7CB at decreasing temperatures (from top to bottom) The $N_x$-N transition is noted and occurs at a reduced temperature of 0.9640.

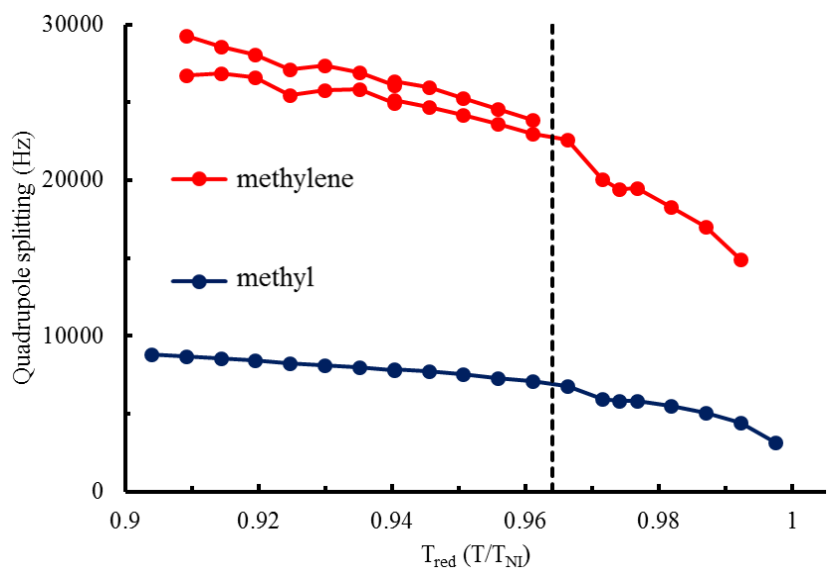

**Figure SI.6.** The temperature dependence (in reduced temperature) of the quadrupole splitting of the methyl and its adjacent methylene group of decane-$d_{22}$ in CB7CB with the N to Nx phase transition shown as dotted line.

*Order Parameters of the cyanobiphenyl mesogenic cores*

Using the same sample (CB11CB with 2 wt% decane-$d_{22}$), $^{13}$C NMR spectra were recorded at 90.6 MHz. The sample was introduced into the spectrometer, and after equilibration to the desired temperature, 1000 scans were averaged with a spectral width of 36 kHz. An inverse gated pulse sequence ($^1$H decoupled during acquisition time) was used with a 90° pulse width of 11 µs and 2 s recycle time. The same procedure was used for CB9CB, CB7CB, and the monomer, 5CB.

$^{13}$C spectra of CB11CB at decreasing temperature with $^1$H decoupling is shown in Figure 9. In the oriented phase, the chemical shift anisotropy (CSA) of the cyano (CN) group causes it to overlap with the chemical shift of the aliphatic carbons. In order to identify the CN group, a $^1$H coupled experiment was performed (Figure 10). The CN peak does not have any observable dipolar coupling in the $^1$H coupled experiment, so its chemical shift can be distinguished from the aliphatic peaks.

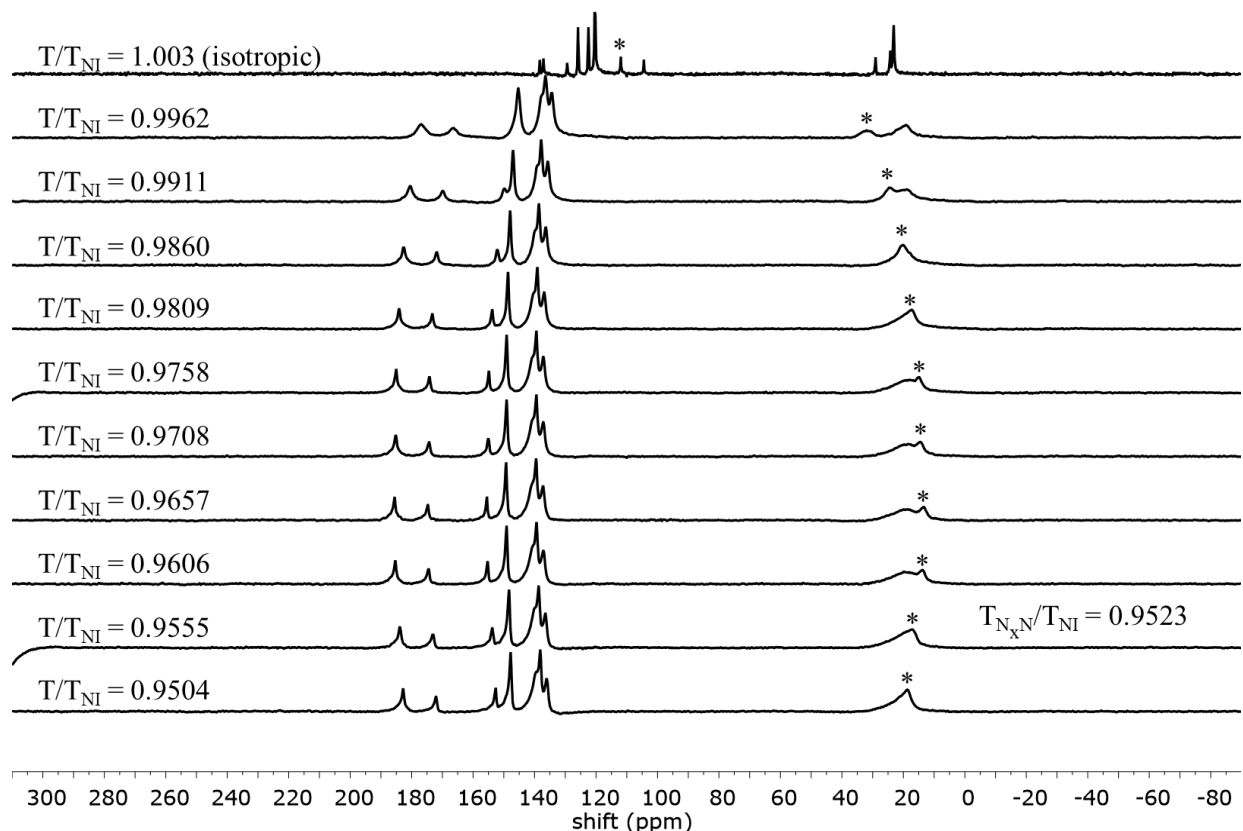

**Figure SI.7.** $^{13}$C spectra of CB11CB at decreasing temperature with $^1$H decoupling. The first spectrum is in the isotropic phase. The cyano group (*) is identified.

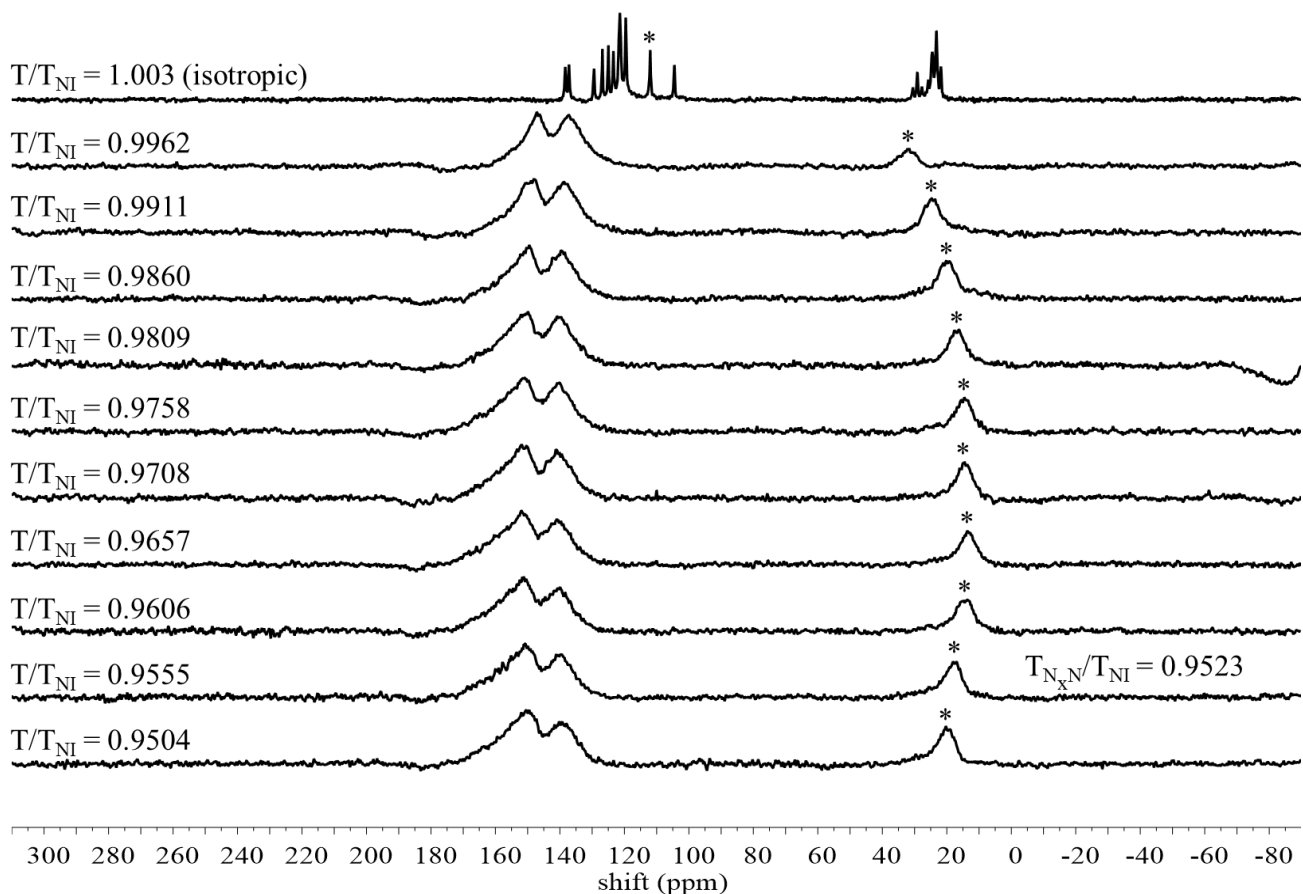

**Figure SI.8.** $^{13}$C spectra of CB11CB at decreasing temperature without $^1$H decoupling. The first spectrum is in the isotropic phase. The cyano group (*) is identified.

The temperature dependence of the CN order parameter, $S_{CN}$, is calculated using the observed chemical shifts in equations below.

The observed chemical shift ($\delta_{CN}^{obs}$) at each temperature and the isotropic chemical shift ($\delta_{CN}^{iso}$) were used to calculate the anisotropic chemical shift ($\delta_{CN}^{aniso}$) in the tables for 13C data with eqn (S.1).

$$\delta_{CN}^{aniso} = \delta_{CN}^{obs} - \delta_{CN}^{iso} \quad . \tag{S.1}$$

Equation (S.2) is used to calculate the order parameter of CN ($S_{CN}$). Values for *a* and *b* are taken from values determined for 7CB. [Guo, W.; Fung, B. M. *J. Chem. Phys.* **1991**, *95*, 3917.]

$$S_{CN} \approx \frac{\delta_{CN}^{aniso} - b}{a} \quad , \tag{S.2}$$

where $a = \dfrac{2\Delta\sigma}{3}$, $\Delta\sigma = [\sigma_{zz} - \dfrac{1}{2}(\sigma_{xx} + \sigma_{yy})]$ and *b* in an empirical constant.

*Splitting of methylene hydrogens for decane-d$_{22}$ in CB7CB and a cholesteric nematic liquid crystal*

For 2 wt% decane-d$_{22}$ is dissolved in a normal cholesteric phase (that of cholesteryl nonanoate), the doubling of the C2 quadrupolar splitting does not occur (Figure SI.9). This is a similar to the result in ref [27] for labeled nonanoic acid-d$_2$ dissolved in cholesteryl nonanoate. [Luz, Z.; Poupko, R.; Samulski, E. *J. Chem. Phys.* **1981**, *74*, 5825.]

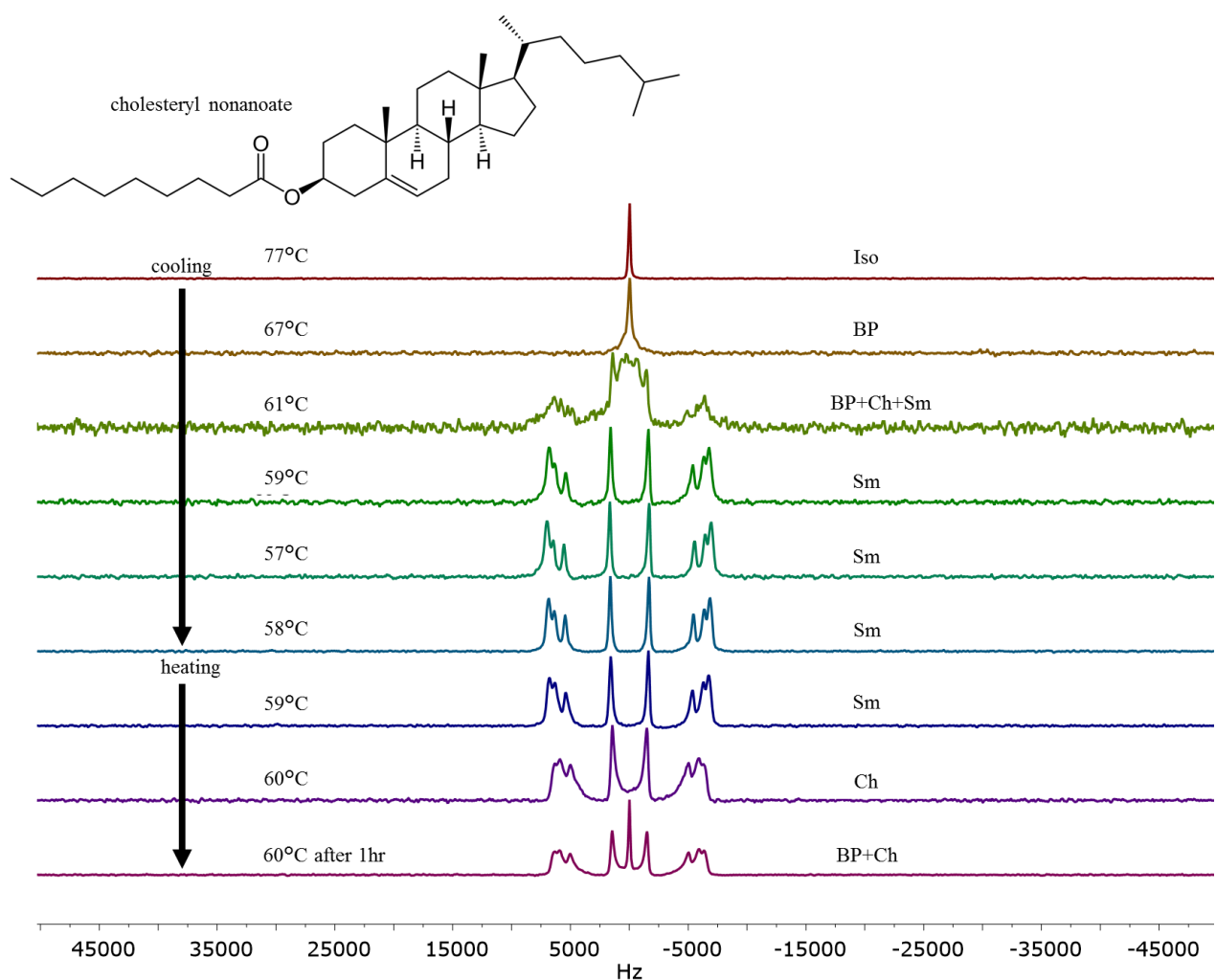

**Figure SI.9.** $^2$H spectra of 2 wt% decane-d22 dissolved in cholesteryl nonanoate cooled into the smectic phase and heated to access the cholesteric phase (spectrum labeled "Ch") as in ref 27.

# Tables of NMR Data
(*deuterium quadrupolar splittings and carbon chemical shifts vs. temperature*)

*Deuterium NMR:*

| 2 wt% decane-$d_{22}$ in CB-C11-CB | | | | $T_{NI}$ (obs) | 392 |
|---|---|---|---|---|---|
| T (K) | $T_{red}$ | C1 (Hz) | C2 (Hz) | | |
| 391.8 | 0.999 | 4201 | 14066 | | |
| 390.6 | 0.996 | 5104 | 17196 | | |
| 386 | 0.985 | 6572 | 21973 | | |
| 383.7 | 0.979 | 7048 | 23500 | | |
| 382.5 | 0.976 | 7258 | 24182 | | |
| 381.3 | 0.973 | 7435 | 24766 | | |
| 380.2 | 0.97 | 7611 | 25201 | | |
| 379 | 0.967 | 7706 | 25543 | | |
| 377.8 | 0.964 | 7788 | 25883 | | |
| 376.7 | 0.961 | 7851 | 25948 | | |
| 375.5 | 0.958 | 7894 | 26092 | | |
| 374.3 | 0.955 | 7979 | 26333 | | |
| 373.2 | 0.952 | 8072 | 26707 | | |
| 372 | 0.949 | 8180 | 26706 | | |
| 370.8 | 0.946 | 8230 | 27031 | | |
| 368.5 | 0.94 | 8411 | | | |

| 2 wt% decane-d$_{22}$ in CB-C9-CB | | | | T$_{NI}$ (obs) | **389** | | |
|---|---|---|---|---|---|---|---|
| T (K) | T$_{red}$ | C1 (Hz) | C2 (Hz) | C2b (Hz) | C3 (Hz) | C4 (Hz) | C5 (Hz) |
| 388 | 0.9974 | 4579.74 | 15546.93 | | 18484.25 | 19740.57 | |
| 386 | 0.9923 | 5328.61 | 17895.14 | | 20852.82 | 22670.15 | |
| 384 | 0.9871 | 5843.07 | 19550.39 | | 22870.06 | 24473.4 | 24976.87 |
| 382 | 0.982 | 6222.18 | 20785.15 | | 24210.77 | 25999.56 | 26469.93 |
| 380 | 0.9769 | 6534.46 | 21746.34 | | 25345.54 | 27252.08 | 27791.56 |
| 378 | 0.9717 | 6794.29 | 22540.24 | | 26257.45 | 28139.07 | 28663.66 |
| 376 | 0.9666 | 7063.16 | 23215.23 | 23479.29 | 27179.66 | 29198.83 | |
| 374 | 0.9614 | 7291.73 | 23758.17 | 24150.2 | 28004.42 | 29958.76 | 30514.98 |
| 372 | 0.9563 | 7530.43 | 24446.45 | 24946.89 | 28776.17 | 30762.83 | 31289.67 |
| 370 | 0.9512 | 7699.14 | 24804.76 | 25574.84 | 29367.26 | 31511.03 | 31922.66 |
| 368 | 0.946 | 7900.5 | 25301.77 | 26214.09 | 30076.41 | 32058.59 | 32517.94 |
| 366 | 0.9409 | 8045.33 | 25572.82 | 26604.04 | 30446.38 | 32667.24 | 33000.88 |
| 365 | 0.9383 | 8128.02 | 26124.36 | 26960.19 | | | |
| 364 | 0.9357 | 8231.5 | 26259.47 | 27352.38 | 31141.04 | 33291.6 | 34040.26 |
| 362 | 0.9306 | 8324.58 | 26662.85 | 27675.04 | | | |
| 360 | 0.9254 | 8503.4 | 27390.67 | 28531.44 | | | |
| 358 | 0.9203 | 8627.68 | 27501.6 | 28803.54 | | | |
| 356 | 0.9152 | 8823.61 | | | | | |

| 2 wt% decane-d$_{22}$ in CB-C7-CB | | | | T$_{NI}$ (obs) | 385 | | |
|---|---|---|---|---|---|---|---|
| T (K) | T$_{red}$ | C1 (Hz) | C2 (Hz) | C2b (Hz) | C3 (Hz) | C4 (Hz) | C5 (Hz) |
| 384 | 0.9974 | 3180.7 | | | | | |
| 382 | 0.9922 | 4412.12 | 14920.83 | | 17764.68 | 18596.14 | |
| 380 | 0.987 | 5052.44 | 16984.18 | | 19986.4 | 21186.82 | |
| 378 | 0.9818 | 5479.71 | 18300.36 | | 21320.53 | 23003.32 | |
| 376 | 0.9766 | 5805.79 | 19490.59 | | 23413.77 | 24193.76 | 24599.86 |
| 375 | 0.974 | 5822.76 | 19426.64 | | 22672.51 | 24326.82 | |
| 374 | 0.9714 | 5982.55 | 20034.43 | | 23312.36 | 25028.48 | |
| 372 | 0.9662 | 6794.01 | 22584.05 | | 26074.21 | 28118.61 | |
| 370 | 0.961 | 7085.6 | 22999.47 | 23854.03 | 27416.37 | 29189.24 | |
| 368 | 0.9558 | 7311.79 | 23628.76 | 24595.34 | 28175.91 | 30066.31 | |
| 366 | 0.9506 | 7528.15 | 24176.87 | 25271.84 | 28879.47 | 30931.18 | |
| 364 | 0.9455 | 7725.35 | 24705.98 | 25939.61 | 29073.13 | 31469.81 | |
| 362 | 0.9403 | 7871.67 | 25144.86 | 26365.09 | 29791.06 | 32019.22 | |
| 362 | 0.9403 | 7784.17 | 24958.5 | 26108.48 | 29767.5 | 31754.51 | |
| 360 | 0.9351 | 7962.5 | 25868.28 | 26924.4 | 30479.22 | 32239.28 | |
| 358 | 0.9299 | 8125.02 | 25772.28 | 27363.5 | 31077.69 | 32793.42 | |
| 356 | 0.9247 | 8241.89 | 25459.49 | 27117.84 | 31243.67 | 33476.35 | |
| 354 | 0.9195 | 8445.74 | 26628.47 | 28049.46 | 31621.89 | 33277.44 | 34091.95 |
| 352 | 0.9143 | 8555.4 | 26831.47 | 28590.38 | 32354.64 | 33979.43 | 34528.83 |
| 350 | 0.9091 | 8696.77 | 26755.16 | 29298.12 | 32031.41 | 34863.05 | |
| 348 | 0.9039 | 8837.28 | | | | | |

| 2 wt% decane-d$_{22}$ in CB-C10-CB | | | | T$_{NI}$ (obs) | 425 | |
|---|---|---|---|---|---|---|
| T (K) | T$_{red}$ | C1 (Hz) | C2 (Hz) | C3 (Hz) | C4 (Hz) | C5 (Hz) |
| 420 | 0.98824 | 13694.1 | 45126.3 | 52243.3 | 56081.2 | 57080.9 |
| 415 | 0.97647 | 14463.2 | 47542.1 | 55075.1 | 58687.4 | 59508.2 |
| 410 | 0.96471 | 15147.3 | 49433.8 | 57265.5 | 60844.3 | 61737.9 |
| 405 | 0.95294 | 15827.9 | 51322.1 | 58759.6 | 63113.7 | 63910.9 |
| 400 | 0.94118 | 16380.5 | 52620 | 60411.7 | 64418.4 | 64920.9 |

| 2 wt% decane-d$_{22}$ in 5CB | | | T$_{NI}$ (obs) | 305 | |
|---|---|---|---|---|---|
| T (K) | T$_{red}$ | C1 (Hz) | C2 (Hz) | C3 (Hz) | C4 (Hz) | C5 (Hz) |
| 296.6 | 0.97246 | 7411.49 | 24318.5 | 28588.8 | 30977.1 | 31737.5 |
| 298 | 0.97705 | 7087.32 | 23255.7 | 27375.3 | 29560.3 | 30498 |
| 299 | 0.98033 | 6785.75 | 22292.3 | 26248.4 | 28372.1 | 29221.1 |
| 300 | 0.98361 | 6489.09 | 21338.8 | 25116.5 | 27178.6 | 27983.6 |
| 301 | 0.98689 | 6110.42 | 20017.5 | 23580.7 | 25740.9 | 26082.4 |
| 302 | 0.99016 | 5689.61 | 18695.6 | 22004.7 | 24016.7 | 24483.3 |
| 303 | 0.99344 | 5084.33 | 16787 | 19779.1 | 21423.3 | 22034.3 |
| 304 | 0.99672 | 4326.71 | 14191 | 16755.1 | | 18518.2 |

*Carbon NMR:*

| CB-C11-CB | | T$_{NI}$ (obs) | 393.5 | | |
|---|---|---|---|---|---|
| T (K) | T$_{red}$ | $\delta_{CN}^{obs}$ | $\delta_{CN}^{iso}$ | $\delta_{CN}^{aniso}$ | S$_{CN}$ |
| 393 | 0.9987 | 35.16 | 111.96 | -76.8 | 0.3397 |
| 392 | 0.9962 | 29.69 | 111.96 | -82.27 | 0.3687 |
| 391 | 0.9936 | 26.7 | 111.96 | -85.26 | 0.3845 |
| 390 | 0.9911 | 22.88 | 111.96 | -89.08 | 0.4048 |
| 388 | 0.986 | 19.2 | 111.96 | -92.76 | 0.4243 |
| 386 | 0.9809 | 16.43 | 111.96 | -95.53 | 0.439 |
| 384 | 0.9759 | 14.45 | 111.96 | -97.51 | 0.4494 |
| 382 | 0.9708 | 13.3 | 111.96 | -98.66 | 0.4555 |
| 380 | 0.9657 | 13.31 | 111.96 | -98.65 | 0.4555 |
| 379 | 0.9632 | 14.02 | 111.96 | -97.94 | 0.4517 |
| 378 | 0.9606 | 15.91 | 111.96 | -96.05 | 0.4417 |
| 377 | 0.9581 | 16.97 | 111.96 | -94.99 | 0.4361 |
| 376 | 0.9555 | 18.08 | 111.96 | -93.88 | 0.4302 |
| 375 | 0.953 | 19.2 | 111.96 | -92.76 | 0.4243 |
| 374 | 0.9504 | 19.27 | 111.96 | -92.69 | 0.4239 |
| 373 | 0.9479 | 20.48 | 111.96 | -91.48 | 0.41749 |
| 372 | 0.94536 | 20.67 | 111.96 | -91.29 | 0.41648 |
| 371 | 0.94282 | 21.38 | 111.96 | -90.58 | 0.41272 |
| 370 | 0.94028 | 21.02 | 111.96 | -90.94 | 0.41463 |

| CB-C9-CB | | $T_{NI}$ (obs) | 391 | | |
|---|---|---|---|---|---|
| T (K) | $T_{red}$ | $\delta_{CN}^{obs}$ | $\delta_{CN}^{iso}$ | $\delta_{CN}^{aniso}$ | $S_{CN}$ |
| 390 | 0.9974 | 42.7 | 112.25 | -69.6 | 0.3013 |
| 388 | 0.9923 | 35.74 | 112.25 | -76.5 | 0.3382 |
| 386 | 0.9872 | 32.01 | 112.25 | -80.2 | 0.3579 |
| 384 | 0.9821 | 28.88 | 112.25 | -83.4 | 0.3745 |
| 382 | 0.977 | 27.61 | 112.25 | -84.6 | 0.3812 |
| 380 | 0.9719 | 26.72 | 112.25 | -85.5 | 0.386 |
| 378 | 0.9668 | 26.43 | 112.25 | -85.8 | 0.3875 |
| 376 | 0.9616 | 26.88 | 112.25 | -85.4 | 0.3851 |
| 375 | 0.9591 | 27.14 | 112.25 | -85.1 | 0.3837 |
| 374 | 0.9565 | 27.4 | 112.25 | -84.9 | 0.3824 |
| 373 | 0.954 | 27.78 | 112.25 | -84.5 | 0.3803 |
| 372 | 0.9514 | 27.81 | 112.25 | -84.4 | 0.3802 |
| 371 | 0.9488 | 27.9 | 112.25 | -84.4 | 0.3797 |
| 370 | 0.9463 | 28.12 | 112.25 | -84.1 | 0.3785 |
| 369 | 0.9437 | 28.82 | 112.25 | -83.4 | 0.3748 |
| 368 | 0.9412 | 28.29 | 112.25 | -84 | 0.3776 |
| 366 | 0.9361 | 29.12 | 112.25 | -83.1 | 0.3732 |
| 364 | 0.9309 | 29.16 | 112.25 | -83.1 | 0.373 |
| 362 | 0.9258 | 29.77 | 112.25 | -82.5 | 0.3698 |
| 360 | 0.9207 | 30.73 | 112.25 | -81.5 | 0.3647 |
| 358 | 0.9156 | 30.76 | 112.25 | -81.5 | 0.3645 |
| 356 | 0.9105 | 30.2 | 112.25 | -82.1 | 0.3675 |
| 354 | 0.9054 | 30.45 | 112.25 | -81.8 | 0.3662 |
| 352 | 0.9003 | 31.14 | 112.25 | -81.1 | 0.3625 |
| 350 | 0.8951 | 30.84 | 112.25 | -81.4 | 0.3641 |

| CB-C7-CB | | $T_{NI}$ (obs) | 388 | | |
|---|---|---|---|---|---|
| T (K) | $T_{red}$ | $\delta_{CN}^{obs}$ | $\delta_{CN}^{iso}$ | $\delta_{CN}^{aniso}$ | $S_{CN}$ |
| 386 | 0.9948 | 56.83 | 112.04 | -55.21 | 0.2253 |
| 384 | 0.9897 | 46.85 | 112.04 | -65.19 | 0.2782 |
| 382 | 0.9845 | 42.17 | 112.04 | -69.87 | 0.303 |
| 380 | 0.9794 | 39.07 | 112.04 | -72.97 | 0.3194 |
| 379 | 0.9768 | 37.96 | 112.04 | -74.08 | 0.3253 |
| 378 | 0.9742 | 37.13 | 112.04 | -74.91 | 0.3297 |
| 377 | 0.9716 | 36.32 | 112.04 | -75.72 | 0.334 |
| 376 | 0.9691 | 35.59 | 112.04 | -76.45 | 0.3378 |
| 375 | 0.9665 | 33.2 | 112.04 | -78.84 | 0.3505 |
| 374 | 0.9639 | 33.04 | 112.04 | -79 | 0.3514 |
| 373 | 0.9613 | 32.83 | 112.04 | -79.21 | 0.3525 |
| 372 | 0.9588 | 32.86 | 112.04 | -79.18 | 0.3523 |
| 370 | 0.9536 | 33.01 | 112.04 | -79.03 | 0.3515 |
| 368 | 0.9485 | 32.65 | 112.04 | -79.39 | 0.3534 |
| 366 | 0.9433 | 33.04 | 112.04 | -79 | 0.3514 |
| 364 | 0.9381 | 33.17 | 112.04 | -78.87 | 0.3507 |
| 360 | 0.9278 | 34.18 | 112.04 | -77.86 | 0.3453 |
| 358 | 0.9227 | 34.23 | 112.04 | -77.81 | 0.345 |
| 356 | 0.9175 | 34.99 | 112.04 | -77.05 | 0.341 |

| 5-CB | | $T_{NI}$ (obs) | 305 | | |
|---|---|---|---|---|---|
| T (K) | $T_{red}$ | $\delta_{CN}^{obs}$ | $\delta_{CN}^{iso}$ | $\delta_{CN}^{aniso}$ | $S_{CN}$ |
| 297 | 0.9738 | 5.02 | 112.01 | -106.99 | 0.4997 |
| 298 | 0.977 | 7.86 | 112.01 | -104.15 | 0.4846 |
| 299 | 0.9803 | 10.81 | 112.01 | -101.2 | 0.469 |
| 300 | 0.9836 | 13.86 | 112.01 | -98.15 | 0.4528 |
| 301 | 0.9869 | 18.65 | 112.01 | -93.36 | 0.4275 |
| 302 | 0.9902 | 22.71 | 112.01 | -89.3 | 0.4059 |
| 303 | 0.9934 | 29.56 | 112.01 | -82.45 | 0.3696 |
| 304 | 0.9967 | 38.94 | 112.01 | -73.07 | 0.3199 |